# Domain Adaptation in Structural Health Monitoring of Civil Infrastructure: A Systematic Review

Yifeng Zhang, and Xiao Liang

Zachry Department of Civil and Environmental Engineering, Texas A&M University, College Station, TX 77843

## Abstract

This study provides a comprehensive review of domain adaptation (DA) techniques in vibration-based structural health monitoring (SHM). As data-driven models increasingly support the assessment of civil structures, the persistent challenge of transferring knowledge across varying geometries, materials, and environmental conditions remains a major obstacle. DA offers a systematic approach to mitigate these discrepancies by aligning feature distributions between simulated, laboratory, and field domains while preserving the sensitivity of damage-related information. Drawing on more than sixty representative studies, this paper analyzes the evolution of DA methods for SHM, including statistical alignment, adversarial and subdomain learning, physics-informed adaptation, and generative modeling for simulation-to-real transfer. The review summarizes their contributions and limitations across bridge and building applications, revealing that while DA has improved generalization significantly, key challenges persist: managing domain discrepancy, addressing data scarcity, enhancing model interpretability, and enabling adaptability to multiple sources and time-varying conditions. Future research directions emphasize integrating physical constraints into learning objectives, developing physics-consistent generative frameworks to enhance data realism, establishing interpretable and certifiable DA systems for engineering practice, and advancing multi-source and lifelong adaptation for scalable monitoring. Overall, this review consolidates the methodological foundation of DA for SHM, identifies existing barriers to generalization and trust, and outlines the technological trajectory toward transparent, physics-aware, and adaptive monitoring systems that support the long-term resilience of civil infrastructure.

## 1. Introduction

Civil infrastructures such as bridges, buildings, tunnels, and railways form the backbone of modern life, supporting transportation, safety, and economic development. Over decades of operation, these structures inevitably experience deterioration caused by material aging, corrosion, fatigue, and environmental influences. When coupled with occasional extreme events such as earthquakes, hurricanes, or floods, this gradual degradation can accelerate and lead to severe consequences. Without effective monitoring and maintenance, such deterioration may result in significant financial losses or, in extreme cases, structural collapse. Traditional inspection and nondestructive testing (NDT) techniques have been widely used for decades to evaluate structural condition, including visual inspection, ultrasonic testing, and ground-penetrating radar [1, 2]. While these techniques can detect local damage with high precision, they are often labor-intensive, time-consuming, expensive, and sometimes unsafe for large or hard-to-access structures.

To overcome these limitations, structural health monitoring (SHM) has been established as a powerful framework for continuous or periodic structural assessment using data collected from in-situ sensors [3, 4]. With rapid advances in sensing technology, data acquisition, and wireless communication, SHM systems can now operate over long periods and across large spatial scales. Their primary objective is to detect, locate, and quantify damage in real time, thereby enabling predictive maintenance and extending the service life of civil assets [5].

SHM methods are generally divided into model-based and data-driven categories. Model-based approaches rely on analytical or finite-element (FE) models to describe structural behavior and identify damage by updating physical parameters such as stiffness, damping, or mass [6]. These methods provide a direct link between observed data and mechanical properties, which facilitates interpretation, but they are computationally demanding and highly sensitive to modeling assumptions [7]. In contrast, data-driven approaches learn damage patterns directly from measured responses without the need for explicit analytical models. Early studies used signal processing techniques such as Fourier or wavelet transforms, Hilbert–Huang analysis, and autoregressive modeling to extract damage-sensitive features. These features were then classified using algorithms such as support vector machines or decision trees [8, 9]. Although these approaches performed well in controlled experiments, they required expert-designed features and often failed to generalize across different environmental and operational conditions.

The advent of deep learning (DL) has reshaped data-driven SHM by allowing features to be learned automatically from raw measurements. Convolutional neural networks (CNNs) have achieved strong performance in both vision-based [9-11] and vibration-based [12-20] SHM tasks. Recurrent networks and hybrid CNN–LSTM [21, 22] architectures have been adopted for vibration and impedance data, to better capture temporal dependencies [23]. DL has also enabled the integration of visual, vibration, and acoustic data into unified frameworks, improving the reliability of structural damage diagnosis [24]. Despite these advances, transferring DL models from controlled settings to real-world infrastructures remains a formidable challenge [25].

A major difficulty lies in the scarcity of labeled damage data. Structural failure rarely occurs in a measurable or repeatable way, and most SHM datasets are dominated by healthy states. Unsupervised learning methods [26-39] avoid needing damage labels by learning a baseline of normal behavior and flagging deviations as anomalies. This is more practical in principle, yet unsupervised techniques usually offer little insight into what is damaged or where. As shown in Fig. 1, long-term monitoring further complicates the problem due to sensor malfunctions, missing data, and drift, which introduce inconsistencies [40, 41]. Environmental and operational changes, including temperature variation, humidity, wind, and traffic, can substantially affect measured responses, making it difficult to distinguish genuine damage from normal fluctuations [42, 43]. Many existing structures also lack baseline measurements from their original healthy condition [44]. Consequently, models trained under specific conditions often fail to perform well when applied to new environments or structures.

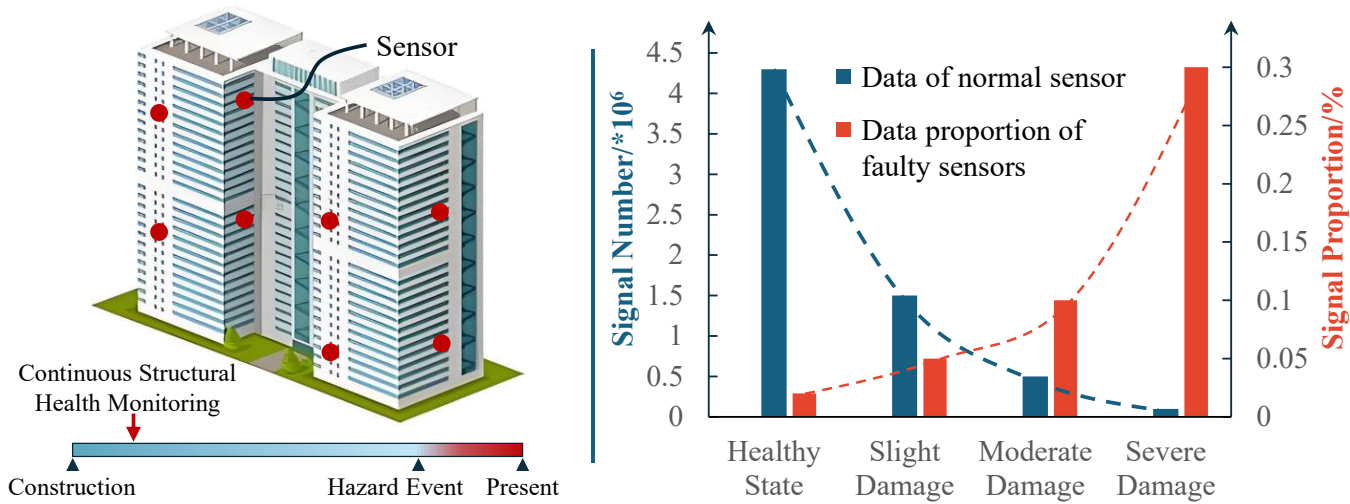

**Fig. 1.** Scarcity of structural damage data.

This performance degradation, commonly referred to as domain shift, occurs when the statistical distribution of training data differs from that of deployment data. In SHM, domain shift arises from multiple sources: environmental variability, operational changes, structural differences, missing sensors, the mismatch between simulated and real data, etc. These discrepancies violate the fundamental assumption that training and testing data are identically distributed, resulting in models that generalize poorly across conditions [45, 46]. Developing strategies to mitigate domain shift is therefore critical for achieving reliable SHM systems.

To address these challenges, researchers have increasingly turned to a specific approach, domain adaptation (DA). DA aims to explicitly minimize the statistical divergence between source and target distributions, enabling a model trained on one domain to perform effectively on another without requiring access to target labels [47, 48]. This strategy is particularly well suited to SHM, where prior knowledge from similar structures, historical records, or numerical simulations can be leveraged to enhance performance in new monitoring scenarios. It is visualized in Fig. 2.

The application of DA in SHM can be broadly organized into two complementary directions: statistical alignment and adversarial alignment. Statistical alignment explicitly seeks to minimize the discrepancy between source and target feature distributions so that models trained on one domain remain valid on another. Early statistical alignment methods began with Kernel Mean Matching (KMM), which reweighted source-domain samples to match the target-domain distribution in the feature space [49]. This idea was later generalized into the Maximum Mean Discrepancy (MMD) framework, where domain alignment is achieved by directly minimizing the distance between the mean embeddings of the two distributions in a reproducing kernel Hilbert space [50]. Building upon this foundation, methods such as Transfer Component Analysis (TCA) introduced mechanisms to reweight or project features into subspaces that preserve intrinsic geometry while reducing cross-domain discrepancy [51]. Later extensions, including Joint Distribution Adaptation (JDA) and Correlation Alignment (CORAL), jointly align both marginal and conditional distributions or match second-order statistics to achieve better class consistency between domains [52, 53]. These techniques have been successfully applied to vibration-based SHM, where environmental or operational variations alter sensor responses, by learning domain-invariant feature spaces that isolate damage-related information from external influence.

Adversarial alignment addresses the same challenge from a different perspective. Instead of measuring distances explicitly, it formulates adaptation as a minimax game between two components: a feature extractor and a domain discriminator. The discriminator learns to identify the domain origin of features, while the extractor updates its parameters to prevent such discrimination, thus forcing the latent representations of source and target data to become indistinguishable. The Domain-Adversarial Neural Network (DANN) framework popularized this principle, later extended by Conditional Adversarial Domain Adaptation (CDAN), which conditions the discriminator on class predictions to preserve category-specific information [54, 55]. These adversarial strategies are particularly useful in SHM applications where structural responses vary nonlinearly across environments or assets. They enable the model to focus on invariant structural signatures, improving generalization without requiring labeled target data.

Recent advances have expanded DA from purely data-driven approaches to hybrid frameworks that incorporate physics or generative synthesis. Physics-informed DA integrates governing principles of structural dynamics into the loss function, ensuring that learned features remain consistent with the underlying physical behavior[56]. Multi-source domain adaptation (MSDA) further generalizes the paradigm by exploiting information from multiple source structures or simulations. Adaptive weighting mechanisms are used to evaluate the similarity of each source to the target, allowing the model to selectively emphasize more relevant domains while suppressing negative transfer [57]. Generative frameworks, such as cycle-consistent adversarial networks, have also been employed to translate simulated responses into realistic target-like data. These models capture both structural and environmental variability, improving the fidelity of simulation-to-real transfer and enhancing model robustness in field deployment [58, 59]. Collectively, these developments demonstrate the evolution of DA in SHM, from simple statistical matching toward comprehensive hybrid frameworks that integrate learning, physics, and data synthesis to achieve domain-invariant and interpretable representations.

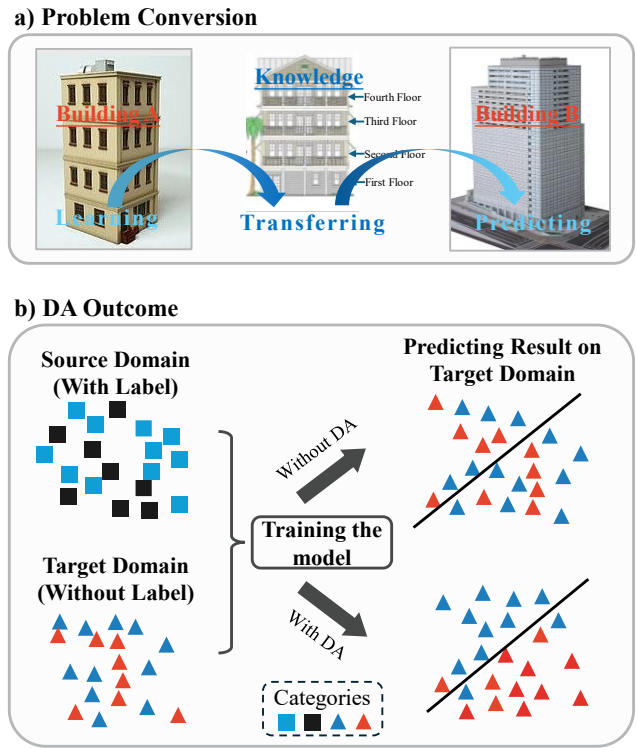

**Fig. 2.** The approach of using DA to address the issue of data scarcity.

Despite the rapid progress of DA in SHM, the existing body of work remains fragmented. Many existing studies focus on specific scenarios, such as temperature compensation or knowledge transfer between two structures, rather than providing a unified and comprehensive framework. Differences in datasets, evaluation metrics, and experimental protocols hinder reproducibility and objective comparison. Moreover, uncertainty quantification and model interpretability are often neglected, even though they are vital for engineering decisions. These gaps underscore the need for systematic reviews that synthesize progress across multiple domains and clarify future research priorities.

This work aims to provide a comprehensive review of recent advances in DA for vibration-based SHM applications in civil engineering. A total of 68 peer-reviewed studies published up to October 2025 were systematically collected and analyzed to summarize the development of unsupervised and transfer-based DA methods for structural damage identification. The review specifically focuses on vibration-based approaches and excludes purely vision-based or image-driven SHM studies, emphasizing algorithms that handle distributional discrepancies, structural heterogeneity, and environmental variability through domain alignment, feature transfer, or physics-informed learning. We tallied the number of publications per year employing two alignment strategies: statistical alignment and adversarial alignment. As shown in Fig. 3, statistical alignment has steadily gained prominence in SHM and its applications are consistently higher than those of adversarial alignment, due to its greater interpretability.

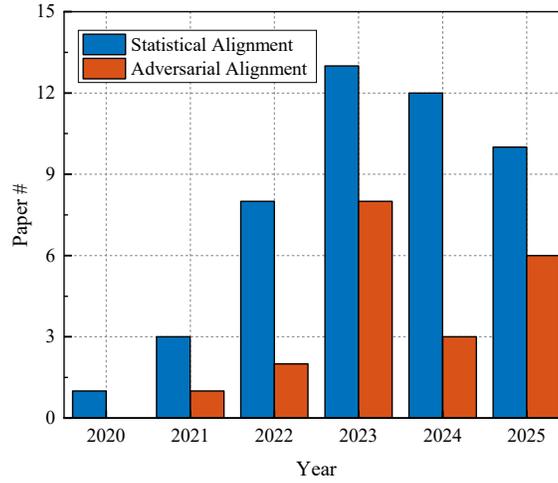

**Fig. 3.** Annual number of SHM studies employing statistical alignment vs. adversarial alignment (by year).

The article selection followed a rigorous and reproducible process. Peer-reviewed publications were retrieved from well-established academic databases, including Web of Science, ScienceDirect, ASCE Library, Wiley Online Library, IEEE Xplore Digital Library, and SAGE. The search was conducted using combinations of at least two of the following keywords: DA, knowledge transfer, domain shift, cross domain, structural health monitoring, structural damage detection, and physics informed. Only vibration-based SHM papers that explicitly applied DA or transfer learning to address distribution mismatch between simulated and real data, between structures, or under varying environmental conditions were included. This selection strategy ensures a focused yet representative overview of the current state-of-the-art in DA-based SHM research. Tables 1 and 2 illustrate the applications of the two alignment strategies, respectively.

**Table 1** Applications of Statistical Alignment DA in Civil SHM

| Authors | Year | Feature data | Algorithm | Test structure | Application sections |
|---|---|---|---|---|---|
| Rezazadeh et al.[60] | 2025 | SHM datasets under environmental variability | Review of baseline compensation and adaptation methods | Civil infrastructure systems | • Environmental variations |
| Ghiasi et al.[61] | 2025 | Drive by vehicle acceleration signals | Unsupervised domain adaptation with statistical alignment | Railway tracks under drive by testing | • Environmental variations • Simulation-to-real transfer |
| Ghiasi et al.[62] | 2025 | Vehicle mounted drive by vibration signals | Progressive distribution alignment with label correction | Railway track structures | • Environmental variations • Simulation-to-real transfer |
| Pinello et al.[63] | 2025 | Ultrasonic guided wave signals | Multilinear principal component analysis with domain adaptation | Ultrasonic plate or pipe specimens | • Sensor configuration |
| Wan et al.[64] | 2025 | Vibration responses of long span bridge under wind | Statistical domain adaptation for pattern recognition | Long span cable supported bridge | • Sensor configuration |

| Authors | Year | Feature data | Algorithm | Test structure | Application sections |
|---------|------|--------------|-----------|----------------|---------------------|
| Zheng et al.[65] | 2025 | Incomplete SHM time series from sensors | Unsupervised domain adaptation with pretrained encoder | Long-term monitored engineering structures | • Sensor configuration |
| Lu et al.[66] | 2025 | Acceleration records from scale bridge model | Convolutional neural network with domain adaptation | Reduced scale cable stayed bridge in laboratory | • Structural differences |
| Ferreira et al.[67] | 2025 | Bridge vibration features and model parameters | Transfer learning with Bayesian calibration | Twin concrete highway bridges | • Structural differences |
| Giglioni et al.[68] | 2025 | Laboratory bridge vibration measurements | Domain adaptation with population-based transfer | Laboratory multi-span continuous girder bridges | • Structural differences |
| Nie et al.[69] | 2025 | Vibration features from multiple structures | Deep transfer learning with domain generalization | Civil structures under simulated and experimental loading | • Structural differences |
| Zhang et al.[70] | 2025 | Structural vibration features for multiple classes | Class weighted subdomain adaptation network | Numerical and experimental structures | • Simulation-to-real transfer |
| Wang et al.[71] | 2025 | Track acceleration signals and reconstructed displacement | Frequency domain invariant representation with domain adaptation | Railway track and vehicle system | • Simulation-to-real transfer |
| Wang et al.[72] | 2025 | Rail and vehicle vibration signals with semantic features | Zero shot domain adaptation with semantic feature learning | Rail fastener systems on railway lines | • Simulation-to-real transfer |
| Xu et al.[73] | 2025 | Simulated and experimental concrete response images | Self-supervised domain adaptation with deep encoders | Concrete beam or panel specimens | • Simulation-to-real transfer |
| Cavanni et al.[74] | 2024 | Vibration features from dome monitoring | Statistical domain adaptation with feature projection | Historical masonry domes | • Environmental variations |
| Souza et al.[75] | 2024 | Bridge vibration features | Joint distribution adaptation | Monitored highway bridges | • Environmental variations |
| Huang et al.[76] | 2024 | Dynamic responses from digital twin models | Feature transferable digital twin with domain adaptation | Truss bridge numerical and monitored models | • Environmental variations |
| Giglioni et al.[77] | 2024 | Bridge modal indicators and vibration features | Transfer component style domain adaptation | Steel concrete composite bridge spans | • Structural differences |

| Authors | Year | Feature data | Algorithm | Test structure | Application sections |
|---------|------|--------------|-----------|----------------|---------------------|
| Marasco et al.[78] | 2024 | Pedestrian bridge vibration measurements | Unsupervised transfer learning with TCA style mapping | Prestressed concrete pedestrian bridges | • Structural differences |
| Giglioni et al.[79] | 2024 | Modal features of adjacent bridge spans | Statistical domain adaptation with feature alignment | Steel concrete composite bridge spans | • Structural differences |
| Giglioni et al.[80] | 2024 | Population level structural features | Domain adaptation for population based SHM | Laboratory and numerical structural populations | • Structural differences |
| Duran et al.[81] | 2024 | Structural response images or time series for CNN | Convolutional neural network with transfer learning | Numerical and experimental civil structures | • Structural differences |
| Xiao et al.[82] | 2024 | Multichannel bridge vibration signals | Multi-channel subdomain adaptation deep transfer learning | Real bridges with multiple damage types | • Structural differences |
| Ghazimoghadam et al.[82] | 2024 | Vibration time series with damage and healthy states | Multi head self-attention LSTM autoencoder with transfer | Laboratory bridge or frame structures | • Environmental variations • Simulation-to-real transfer |
| Song et al.[83] | 2024 | Bridge vibration responses | Dynamic distribution adaptation with transfer learning | Simply supported box girder bridges | • Simulation-to-real transfer |
| Xiong et al.[84] | 2024 | Multi channel seismic response time series | Zero shot domain adaptation with autoencoder and CNN | Seismic benchmark frames and shake table tests | • Simulation-to-real transfer |
| Sawant et al.[85] | 2023 | Ultrasonic guided wave signals | Unsupervised transfer learning with temperature compensation | Laboratory ultrasonic SHM specimen | • Environmental variations |
| Ardani et al.[86] | 2023 | Bridge vibration responses | Proper orthogonal decomposition with transfer learning | Highway bridge with field monitoring | • Environmental variations • Structural differences |
| Pan et al.[87] | 2023 | Sensor time series from SHM systems | Transfer learning based anomaly detection | Long term monitored civil structures | • Sensor configuration |
| Chen et al.[88] | 2023 | Raw vibration time series and time frequency maps | Deep convolutional transfer learning with MMD | Benchmark test structures and bridges | • Structural differences |
| Yano et al.[89] | 2023 | Bridge vibration features before and after retrofit | Joint distribution adaptation with transfer learning | Railway bridges in Europe | • Structural differences |
| Yano et al.[90] | 2023 | Modal features and damage indices | Transfer component analysis with Gaussian process regression | Three story benchmark building simulation and experiment | • Environmental variations• Simulation-to-real transfer |

| Authors | Year | Feature data | Algorithm | Test structure | Application sections |
|---|---|---|---|---|---|
| Bao et al.[91] | 2023 | Vibration signals from simulated and real structures | Deep transfer learning network | Civil structures with limited field data | • Simulation-to-real transfer |
| Figueiredo et al.[92] | 2023 | Simulated and measured bridge vibration data | Transfer learning from numerical models to field data | Bridge structures including Z24 | • Simulation-to-real transfer |
| Martakis et al.[93] | 2023 | Building acceleration responses and modal features | Fusion of damage sensitive features with domain adaptation | Real reinforced concrete buildings under shaking | • Simulation-to-real transfer |
| Yuan et al.[94] | 2023 | Vibration signals from isolators and track slabs | Unsupervised cross domain damage detection with DA | Metro floating slab track system | • Simulation-to-real transfer |
| Xiangyu et al.[95] | 2023 | Seismic response signals of high arch dams | Unsupervised domain adaptation with sparse autoencoder | High arch dam numerical and monitored models | • Simulation-to-real transfer |
| Quqa et al.[96] | 2023 | Electrical impedance tomography conductivity maps | Transfer learning with feature alignment | Concrete specimens with conductive coating | • Simulation-to-real transfer |
| Gardner et al.[97] | 2022 | Vibration features for multiple structures | Gaussian mixture model with domain adaptation | Population of simulated and measured structures | • Environmental variations |
| Poole et al.[98] | 2022 | Generic SHM feature sets | Statistical alignment with distribution matching | Benchmark SHM datasets for structures | • Environmental variations |
| Chamangard et al.[99] | 2022 | Structural vibration records for CNN | Convolutional neural network with transfer learning | Laboratory and numerical civil structures | • Environmental variations |
| Sajedi and Liang[100] | 2022 | Simulated structural responses at candidate sensor locations | Deep generative model with Bayesian optimization | Numerical model of civil structure | • Sensor configuration |
| Xiao et al.[101] | 2022 | Multichannel bridge acceleration signals | Multi channel deep transfer learning with domain adaptation | Real bridges with multiple sensors | • Structural differences |
| Xiao et al.[102] | 2022 | Bridge vibration time series | Subdomain adaptation deep transfer learning | Multiple bridges with unlabeled health states | • Structural differences |
| Tronci et al.[103] | 2022 | Cepstral features and x vector representations | Transfer learning from speaker recognition features | Z24 bridge and related bridge datasets | • Structural differences |
| Zhang et al.[104] | 2022 | Vibration response data and model parameters | Transfer learning with Bayesian updating | Numerical and experimental structural models | • Simulation-to-real transfer |
| Lin et al.[105] | 2022 | Structural dynamic response signals | Dynamics based deep transfer learning | Numerical and experimental frame or beam structures | • Simulation-to-real transfer |

| Authors | Year | Feature data | Algorithm | Test structure | Application sections |
|---------|------|--------------|-----------|----------------|---------------------|
| Silva et al.[106] | 2021 | Impedance measurements | Transfer component analysis | Laboratory impedance based SHM specimen | • Environmental variations |
| Gardner et al.[107] | 2021 | Structural features across heterogeneous systems | Mapping and transfer framework for population based SHM | Heterogeneous structural populations | • Structural differences |
| Ozdagli and Koutsoukos[108] | 2021 | Simulated and measured vibration signals | Domain adversarial style statistical adaptation under uncertainty | Structural testbeds with numerical and experimental data | • Simulation-to-real transfer |
| Gardner et al.[109] | 2020 | Generic SHM feature sets from multiple examples | Statistical domain adaptation with feature mapping | Representative SHM structures and case studies | • Structural differences |

**Table 2** Applications of Adversarial Alignment DA in Civil SHM

| Authors | Year | Feature data | Algorithm / DA strategy | Test structure | Application sections |
|---------|------|--------------|------------------------|----------------|---------------------|
| Talaei et al.[110] | 2025 | Bridge and vehicle dynamic responses | Hybrid adversarial domain adaptation | Prestressed concrete bridge with moving vehicles | • Simulation-to-real transfer |
| Wang et al.[111] | 2025 | Monitoring data and simulated responses from digital twin | Physics informed machine learning with adversarial transfer | Civil structures represented by hybrid digital twin | • Simulation-to-real transfer |
| Zhang et al.[112] | 2025 | Transmissibility functions from vibration data | Adversarial and discrepancy based unsupervised domain adaptation | Buildings and tall tower including Canton Tower | • Simulation-to-real transfer |
| Li et al.[113] | 2024 | Structural vibration time series | Conditional adversarial domain adaptation | Beam and frame structures under dynamic loading | • Simulation-to-real transfer |
| Ge and Sadhu[114] | 2024 | Simulated and field vibration responses | Physics informed generative adversarial learning with self attention | Steel beam and bridge structures | • Simulation-to-real transfer |
| Nie et al.[115] | 2024 | Bridge vibration signals | Adversarial based transfer learning | Bridge structures with multiple damage cases | • Simulation-to-real transfer |
| Liao and Qiao[116] | 2023 | Ultrasonic guided wave signals across scenes | Adversarial style cross scene feature transfer | Reinforced concrete specimens with interfacial debonding | • Sensor configuration |
| Luleci and Catbas[117] | 2023 | Bridge vibration responses under different conditions | Structural state translation with generative mapping | Prestressed concrete highway bridges | • Structural differences |

| Authors | Year | Feature data | Algorithm / DA strategy | Test structure | Application sections |
|---|---|---|---|---|---|
| Liu et al.[118] | 2023 | Drive by vehicle acceleration responses on bridges | Hierarchical multi task adversarial domain adaptation | Multiple highway bridges under drive by tests | • Sensor configuration • Structural differences |
| Xiao et al.[119] | 2023 | Multichannel bridge vibration data | Adversarial auxiliary weighted subdomain adaptation | Real bridges under open set conditions | • Structural differences |
| Wang et al.[120] | 2023 | Lamb wave ultrasonic signals from simulation and experiment | Adversarial domain adaptation between finite element and tests | Metallic plate specimens with fatigue cracks | • Simulation-to-real transfer |
| Luleci et al.[121] | 2023 | Measured undamaged vibration signals | Cycle consistent generative adversarial network | Steel grandstand and structural test stands | • Simulation-to-real transfer |
| Soleimani Babakamali et al.[122] | 2023 | Spectral features of structural responses | Zero shot generative adversarial transfer learning | Z24 bridge and other experimental structures | • Simulation-to-real transfer |
| Luleci and Catbas[123] | 2022 | Structural vibration responses from multiple structures | Domain generalization with state translation model | Civil structures including bridges and frames | • Structural differences |
| Xiao et al.[124] | 2022 | Bridge vibration signals with multiple damage types | Adversarial fuzzy weighted deep transfer learning | Real bridges with multiple damage scenarios | • Structural differences |
| Xu and Noh[125] | 2021 | Seismic response time series from multiple buildings | Physics informed multi source domain adversarial network | Numerical and experimental building structures | • Structural differences |

Section 2 reviews core DA principles and algorithms. Section 3 categorizes DA applications based on domain shift types: environmental variation, sensor configuration, structural differences, and simulation-to-real transfer. Section 4 discusses key challenges and outlines future research priorities. Section 5 concludes with reflections on the path toward scalable, interpretable, and adaptive SHM systems.

## 2. Domain adaptation concepts and strategies

This section presents the theoretical and algorithmic foundations of DA methods used in SHM. It bridges the gap between general machine learning principles and their application to civil infrastructures under changing environmental and operational conditions. The following subsections introduces the motivation for employing DA in SHM and further summarizes the fundamental principles of DA, including statistical alignment, adversarial learning, and it reviews representative algorithms and their evolution from shallow subspace methods to modern deep adaptation networks.

### 2.1 Motivations

In SHM, the central challenge lies not only in the scarcity of labeled data but also in the fact that the statistical properties of signals change across structures, environments, and time. This continual variation creates a learning problem that cannot be solved by conventional supervised training on a single dataset. DA directly addresses this issue by learning representations that remain valid when the training (source) and deployment (target) domains differ [126]. Its goal is to minimize the degradation in model performance caused by such shifts, allowing a detector or classifier trained on one structure or environmental condition to generalize effectively to another without requiring extensive re-labeling.

This motivation is grounded in theory. Generalization bounds for domain shift show that the expected target error can be decomposed into three parts: the source error, a divergence term measuring the discrepancy between source and target distributions, and a compatibility term describing how similar their labeling functions are [126]. When this divergence is large, collecting more source data does not improve performance; the learner must instead reduce the divergence by learning domain-invariant features. DA achieves this by transforming or reweighting feature spaces so that source and target data become statistically aligned while preserving task-relevant information [127]. In doing so, it provides a principled mechanism to control target error even in the presence of strong distribution mismatch.

A second motivation arises from the diversity of shifts encountered in SHM. Changes in the input distribution while the labeling rule remains stable motivate importance-weighting and reweighting strategies that correct for sampling bias between domains [128, 129]. More complex shifts require calibration and model selection procedures that remain stable when the proportion of healthy and damaged states changes across assets or seasons [130]. When multiple sources are available, the learner must identify and emphasize those most similar to the target, since negative transfer occurs when distant or inconsistent sources dominate the adaptation process.

A third motivation concerns robustness and interpretability. SHM systems must remain reliable across long monitoring periods and under diverse operational conditions. DA encourages representations that are invariant to benign environmental fluctuations yet sensitive to genuine structural changes. Recent theoretical work on invariant risk minimization formalizes this principle by seeking predictors whose performance remains stable across varying environments, precisely the property required to distinguish environmental effects from structural damage [131]. Geometric formulations of distribution shift further enable quantitative and auditable reasoning about domain alignment quality [132]. Collectively, these perspectives justify the use of DA as a foundation for scalable, interpretable, and robust SHM systems capable of generalizing across assets, conditions, and time with controlled generalization risk.

## 2.2 Domain adaptation strategies

DA aims to reduce the distributional discrepancy between source and target domains so that models trained on labeled source data can perform reliably on unlabeled or sparsely labeled target data. In SHM, this capability is essential because measurements obtained from different structures, environmental conditions, or sensor configurations often exhibit distinct statistical characteristics. Such discrepancies cause supervised models to lose predictive accuracy when deployed outside their original domain.

DA provides a mechanism for bridging this gap by learning representations that are invariant to domain-specific variations yet remain sensitive to structural changes. Through appropriate alignment of data distributions or model responses, DA enables knowledge transfer across diverse monitoring scenarios without extensive relabeling. Two principal directions are generally recognized: statistical alignment, which seeks to minimize a measurable distance between domains in a latent space, and adversarial alignment, which encourages domain invariance through discriminative training. These complementary approaches form the foundation for robust and transferable SHM frameworks.

### 2.2.1 Statistical alignment

Statistical alignment methods aim to minimize the distributional discrepancy between source and target domains by constructing a latent feature space where representations align according to defined criteria [133]. As conceptualized in Fig. 4, methods relying on statistical alignment seek to find optimal feature mappings for both domains, thus enabling a single classifier to effectively separate classes across the new, unified feature space. A foundational technique in this family is KMM, where weights are assigned to source samples such that their weighted mean in a reproducing-kernel Hilbert space matches that of the target. In effect, KMM implements a reweighting scheme that enforces a match of means across domains [134].

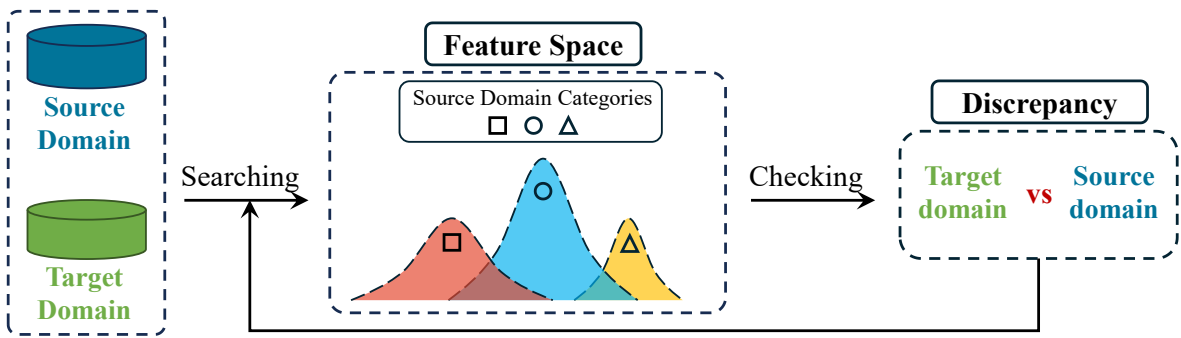

**Fig. 4.** Statistical alignment in domain adaptation.

From this idea emerges the principle of MMD, which generalizes KMM by directly learning a feature transformation that reduces the distance between mean embeddings of transformed source and target distributions. This mapping aligns marginal distributions in the latent space without requiring target labels [135]. To extend alignment to class-conditional structure, alignment can include additional constraints on the joint or conditional distributions, for example by integrating label information or pseudo-labels into the measure, thus preserving consistency across classes [53].

Beyond mean embedding, methods may match higher-order statistics or subsets of distributional structure. For instance, covariance alignment aligns second-order statistics, while some approaches consider higher central moments to capture skewness or kurtosis. Another perspective treats each domain as residing on a subspace or manifold and seeks to reduce their distance via projections or interpolated paths on the manifold. A further generalization is optimal transport, which views alignment as transporting one distribution into another in a cost-minimizing way that preserves neighborhood relationships [136].

These alignment strategies require no target labels and often integrate well with standard learning pipelines. They tend to be efficient and interpretable, making them practical baselines in SHM applications where domain shifts are smooth or dominated by low-order effects. Their limitation becomes apparent when shifts are complex, nonlinear, or multimodal, in which case adversarial or hybrid strategies are better suited.

### 2.2.2 Adversarial alignment

Adversarial alignment formulates DA as a game between feature extraction and discrimination. The core components and learning mechanism of Adversarial Domain Adaptation are visualized in Fig. 5. A feature extractor learns representations from input data, a task classifier guides feature learning toward solving the source-label task, and a domain discriminator attempts to distinguish whether a given feature comes from the source or the target domain. During training, the feature extractor is simultaneously optimized to minimize task loss and maximize the discriminator's error. The result is a representation space where source and target are indistinguishable to the discriminator while preserving task-relevant variation [137].

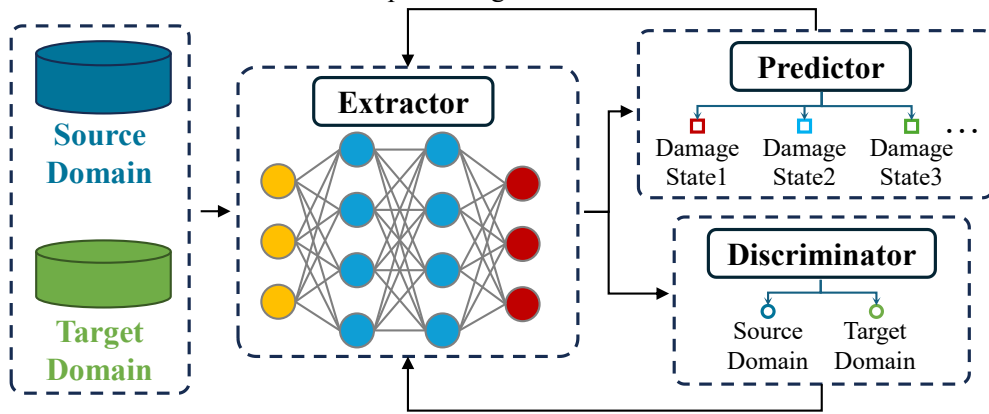

**Fig. 5.** Adversarial alignment in domain adaptation.

In theoretical terms, the discriminator's outputs approximate a divergence measure between domains. Training the extractor to fool the discriminator implicitly minimizes this divergence, often corresponding to criteria such as Jensen–Shannon or Wasserstein distances depending on the discriminator's formulation. Extensions condition the discriminator on task signals or label predictions to focus alignment within decision-relevant regions, reducing the risk of aligning features that belong to different classes [138].

In SHM contexts, the strength of adversarial alignment lies in its capacity to handle complex, nonlinear domain shifts that are hard to capture by static statistical criteria. When environmental factors, sensor changes, and structural differences interact, adversarial methods adaptively mold the feature space. However, achieving a balance is critical: overly strong invariance may mask subtle damage indicators, so regularization, early stopping, and careful loss weighting are necessary to maintain diagnostic sensitivity.

### 2.3 Domain adaptation algorithms in SHM

The development of DA has followed a clear trajectory from shallow statistical projection to deep adversarial frameworks. Each algorithm embodies a distinct mechanism for reducing domain discrepancy and has been adapted for various SHM contexts, from vibration data to visual inspection.

- **Transfer Component Analysis (TCA)** is one of the earliest statistical DA methods that aims to learn a latent feature subspace where the distributions of the source and target domains become similar while preserving the intrinsic structure of the original data. It extends the principle of MMD by incorporating it into a regularized dimensionality-reduction framework. The principle of MMD is visualized in Fig. 6. TCA seeks a linear or kernelized transformation matrix that maps both domains into a common latent space where the marginal distribution difference between domains is minimized, measured by MMD, and at the same time, data variance is maintained through a reconstruction or smoothness constraint. The optimization problem is typically formulated as minimizing the MMD-based discrepancy term with a regularization on the projection matrix, subject to orthogonality or normalization constraints, which leads to a generalized eigenvalue problem. By applying the learned transformation, TCA enables a classifier trained on the transformed source data to generalize better to the transformed target data. TCA focuses only on aligning the marginal distribution and does not explicitly consider conditional alignment, which makes it simpler but potentially less effective when the class-conditional structures differ. Nevertheless, TCA remains widely used as a baseline due to its mathematical clarity, non-parametric formulation, and flexibility in adopting kernel mappings to capture nonlinear domain shifts [139].

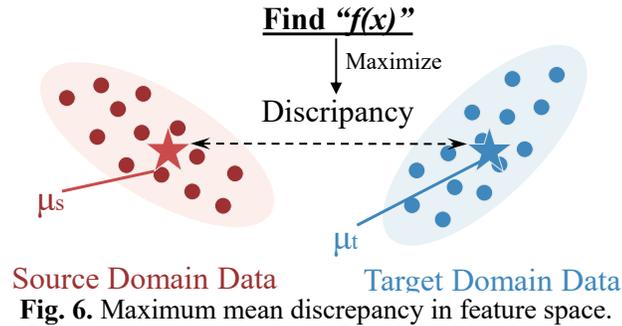

**Fig. 6.** Maximum mean discrepancy in feature space.

- **Joint Distribution Adaptation (JDA)** is a classic unsupervised DA method that learns a feature transformation to reduce distribution mismatch between a labeled source domain and an unlabeled target domain by aligning both the marginal distribution P(X) and the conditional distribution P(Y|X). Building on MMD, JDA formulates an objective that penalizes discrepancies in overall feature means across domains and, simultaneously, class-specific discrepancies computed using pseudo-labels for target samples; the two parts are balanced by a trade-off parameter and solved together with a variance-preserving regularizer through a generalized eigenvalue problem, often in a kernelized space to capture nonlinear structure. In practice, JDA proceeds iteratively: it projects data into a low-dimensional subspace, trains a classifier on the transformed source data, assigns pseudo-labels to target data, recomputes the class-wise MMD terms, and updates the projection until convergence or a fixed number of rounds; common classifiers include 1-NN, SVM, or softmax, and key hyperparameters are the subspace dimension and the marginal–conditional weighting. By explicitly coupling marginal and conditional alignment, JDA typically improves transfer when decision boundaries are approximately shared and the label space is consistent across domains, and it can be enhanced with class-balance weights or structure-preserving terms to stabilize learning. However, its performance can degrade under severe class imbalance, heavy label-shift, or highly noisy pseudo-labels, and the iterative procedure may propagate early mistakes; compared with deep adversarial approaches, JDA is lightweight and transparent but may underfit complex shifts unless kernelization and careful hyperparameter tuning are used [53].

- **Geodesic Flow Kernel (GFK)** is a classic geometric DA method that models the domain shift as a smooth transition between the source and target feature subspaces on a Grassmann manifold. The key idea is that instead of directly aligning the two subspaces, GFK constructs an infinite number of intermediate subspaces lying along the geodesic path that connects the source and target subspaces, and then integrates the contributions from all these subspaces to form a kernel that captures the relationship between domains. Formally, source and target data are first represented in low-dimensional subspaces obtained by principal component analysis (PCA). These subspaces are treated as points on the Grassmann manifold, and a geodesic flow is defined between them to describe continuous domain evolution. The GFK then computes an integral of the inner products of features projected along this flow, leading to a positive semi-definite kernel matrix that measures feature similarity while smoothly bridging the two domains. This kernel can be directly used with standard classifiers such as SVMs to perform transfer learning. By modeling domain shift geometrically rather than statistically, GFK effectively captures gradual variations in feature distributions, especially when the source and target domains share similar global structures. However, since GFK assumes linear subspaces and relies on unsupervised feature extraction, it may struggle with large nonlinear discrepancies or significant class-conditional mismatches, which later motivated nonlinear extensions and deep manifold alignment methods. [140]

- **Subspace Alignment (SA)** is a simple yet effective DA method that aims to align the feature subspaces of the source and target domains through a linear mapping. The main idea is that both domains can be represented by low-dimensional subspaces obtained via PCA, and instead of transforming data into a common space or computing a complex manifold distance, SA directly learns a linear transformation that aligns the source subspace basis to the target subspace basis. Fig. 7 briefly illustrates the concept of SA. Mathematically, this is achieved by finding a transformation matrix that minimizes the Frobenius norm between the transformed source basis and the target basis, resulting in a closed-form solution with very low computational cost. Once the alignment matrix is obtained, source data are projected into the aligned subspace, and a standard classifier trained on the transformed source data can be applied to the target domain. SA is attractive because it is parameter-free, efficient, and easy to implement, while still achieving competitive performance under moderate domain shifts. However, its linear assumption limits its ability to handle nonlinear discrepancies, and it focuses only on global subspace alignment without explicitly addressing class-conditional differences or preserving discriminative information. As a result, SA performs best when the source and target domains share similar feature structures but tends to degrade in more complex or multimodal adaptation scenarios[141].

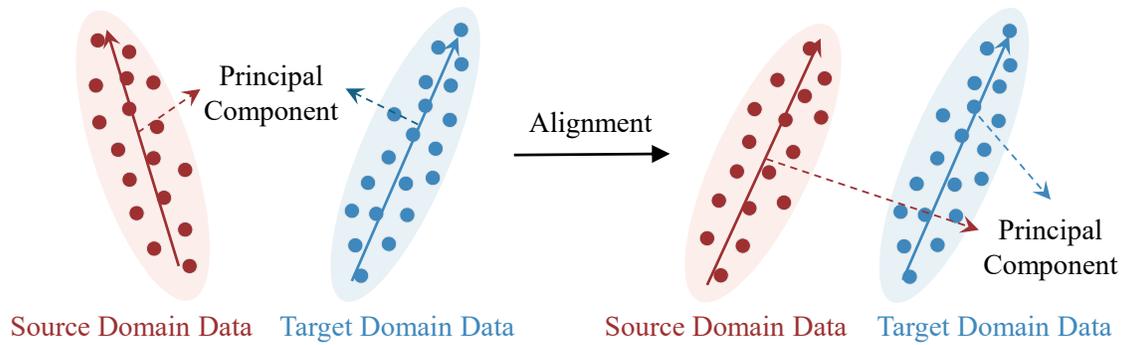

**Fig. 7.** Subspace alignment.

- **Deep Adaptation Network (DAN)** is a DL-based DA framework that integrates distribution alignment into the hidden layers of a neural network to learn transferable feature representations. It extends the idea of kernel-based statistical adaptation, such as in TCA and JDA, to the deep feature space by minimizing the discrepancy between the source and target feature distributions through the MMD criterion. In DAN, multiple layers of a pretrained deep network are adapted by adding MMD loss terms that measure the distance between the mean embeddings of source and target features in a reproducing kernel Hilbert space (RKHS). This multi-layer adaptation enables the model to align not only low-level but also high-level abstract features, improving the generalization ability of the learned representation. The total objective combines the standard classification loss on labeled source samples and the MMD-based adaptation loss, optimized jointly via stochastic gradient descent. To further enhance flexibility, DAN employs multi-kernel MMD, which integrates multiple Gaussian kernels with different bandwidths to capture complex, multi-scale domain shifts. As a result, DAN achieves effective feature alignment without requiring target labels, and it significantly outperforms shallow statistical methods under moderate nonlinear shifts. However, DAN still assumes that source and target share the same label space and may suffer from limitations when class-conditional distributions differ greatly, motivating later extensions such as JAN and DANN that incorporate conditional alignment and adversarial training mechanisms. [142].

- **Joint Adaptation Network (JAN)** is an extension of DAN that aims to align both the marginal and conditional distributions between the source and target domains in deep feature space. While DAN minimizes the discrepancy of feature representations only in terms of their marginal distributions, JAN further considers the joint distribution of the network's activations across multiple layers thereby capturing the relationship between learned representations and their associated class predictions. Specifically, JAN introduces the concept of Joint Maximum Mean Discrepancy (JMMD), which measures the distance between the joint distributions of multi-layer activations from the source and target domains within a reproducing kernel Hilbert space. This joint alignment ensures that samples from both domains with the same underlying class are mapped close to each other, leading to better class-consistent transfer. The network is trained by minimizing a combination of the source classification loss and the JMMD-based alignment loss through stochastic gradient descent. Compared with DAN, which focuses only on marginal feature alignment, JAN provides a more comprehensive adaptation by explicitly coupling features and labels, thereby improving performance under scenarios with conditional shift. Nonetheless, as JAN relies on pseudo-labels for the unlabeled target domain, its effectiveness depends on the accuracy of these estimates, and its performance can degrade under severe label noise or significant domain divergence [143].

- **Deep Correlation Alignment (Deep CORAL)** is a deep DA method that seeks to minimize the domain shift by aligning the second-order statistics of feature representations from the source and target domains. Unlike methods based on MMD that rely on kernel distance in RKHS, Deep CORAL adopts a simpler and computationally efficient approach by directly enforcing the covariance alignment in the feature space of a deep neural network. The method introduces a CORAL loss term, which measures the Frobenius norm of the difference between the covariance matrices of source and target features extracted from one or more layers. During training, this CORAL loss is combined with the standard classification loss on the labeled source data and jointly optimized using backpropagation. By aligning the covariance structure, Deep CORAL ensures that not only the mean but also the feature correlations are similar across domains, encouraging the network to learn domain-invariant representations without requiring any target labels. Compared with adversarial or MMD-based approaches, Deep CORAL is more lightweight, parameter-free, and easy to integrate into existing deep architectures. However, its linear alignment assumption may limit performance when dealing with highly nonlinear or multimodal domain shifts, and it aligns only global statistics without explicitly considering class-level alignment. Despite these limitations, Deep CORAL is widely used as a baseline for efficient and interpretable deep DA [52].

- **Optimal Transport (OT)**–based DA methods formulate domain alignment as a problem of finding the most cost-efficient way to transport the probability mass of the source distribution to match that of the target distribution. Instead of focusing on aligning statistical moments or subspaces, OT provides a principled geometric framework that measures the discrepancy between two distributions by the minimal transport cost required to transform one into the other. Mathematically, this is achieved by solving the Monge–Kantorovich problem, where a transport plan minimizes the expected distance between source and target samples under certain regularization constraints. In practice, the discrete OT problem is often solved in a computationally tractable way by introducing entropic regularization, which leads to the Sinkhorn algorithm for efficient optimization. Once the optimal transport plan is obtained, source samples can be linearly or barycentrically mapped toward target samples, thereby reducing domain shift. OT-based adaptation can be applied either at the feature level or directly in the sample space, and it can be integrated into deep networks as a differentiable loss. Compared with MMD or subspace methods, OT provides a more flexible and theoretically grounded way to match complex, non-overlapping distributions, as it explicitly models sample-wise correspondences. However, OT methods may be computationally expensive for large datasets, sensitive to hyperparameters such as the regularization weight, and less robust when class labels are not aligned or when label shift exists [136].

- **Domain-Adversarial Neural Network (DANN)** is a deep unsupervised DA framework that aligns feature distributions across domains through adversarial learning. The core idea is to train a feature extractor that produces domain-invariant representations by competing against a domain discriminator. The network architecture consists of three main components: a feature extractor shared by both source and target data, a label predictor trained on labeled source samples, and a domain classifier that distinguishes whether a feature comes from the source or target domain. A gradient reversal layer (GRL) is inserted between the feature extractor and the domain classifier so that during backpropagation, the gradient from the domain classifier is multiplied by a negative constant. This mechanism drives the feature extractor to minimize the label classification loss while maximizing the domain classification loss, effectively making the extracted features indistinguishable across domains. The overall objective thus balances accurate task learning on the source domain with strong domain confusion, leading to domain-invariant feature spaces. DANN has been shown to outperform traditional statistical alignment methods such as MMD-based approaches in many signal-based tasks, as it learns nonlinear and task-dependent feature alignment in an end-to-end manner. However, its performance depends on the stability of adversarial training, the relative weighting of the domain loss, and the assumption of shared label space between domains [137].

- **Adversarial Discriminative Domain Adaptation (ADDA)** is an adversarial DA framework that learns domain-invariant feature representations through a two-stage training process. Unlike DANN, which uses a shared feature extractor for both domains with a gradient reversal layer, ADDA trains separate feature extractors for the source and target domains, allowing greater flexibility in modeling domain-specific characteristics. In the first stage, the source feature extractor and classifier are trained in a supervised manner using labeled source data to learn discriminative task-relevant features. In the second stage, the source network is frozen, and a target feature extractor is trained adversarially against a domain discriminator. The discriminator attempts to distinguish between source and target features, while the target extractor tries to fool it by producing features that match the distribution of the source domain. This adversarial objective is formulated similarly to a Generative Adversarial Network (GAN), but applied in the feature space rather than the data space. By decoupling the training of source and target encoders, ADDA avoids potential conflicts between label classification and domain alignment objectives, leading to more stable optimization. It has been successfully applied to various tasks such as cross-domain image recognition, speech adaptation, and vibration-based SHM, particularly when the source and target domains have distinct low-level characteristics but share similar high-level semantics. However, because ADDA relies on a pre-trained source model and unsupervised adversarial training for the target, its performance may depend heavily on the quality of the source representation and the balance between discriminator and generator training [144].

- **Conditional Domain-Adversarial Network (CDAN)** is an extension of the DANN designed to achieve more effective domain alignment by conditioning the adversarial process on class-prediction information. The architecture and information flow of the CDAN framework are summarized in Fig. 8. While DANN aligns only the marginal feature distributions between source and target domains, CDAN additionally considers the dependency between learned features and their corresponding class predictions, thus aligning the joint distribution H(F, G) rather than only the marginal H(F). To achieve this, CDAN introduces a conditioning mechanism where the domain discriminator takes as input not just the feature representations but also their outer product with the softmax output of the classifier. This operation allows the discriminator to model how features interact with class probabilities and to enforce alignment in a class-aware manner. The adversarial training objective then minimizes the domain classification loss while optimizing the feature extractor and classifier to produce domain-invariant and discriminative features simultaneously. In practice, a multilinear conditioning or randomized approximation is often used to control computational cost. CDAN effectively mitigates negative transfer when class-conditional distributions differ across domains and achieves superior performance in complex adaptation scenarios where simple marginal alignment is insufficient. However, it may still be sensitive to noisy pseudo-labels in the target domain, and performance depends on the reliability of classifier outputs used for conditioning [138].

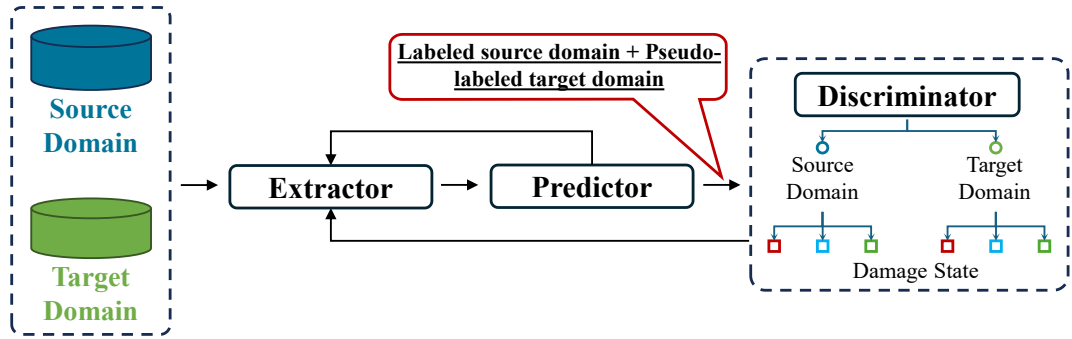

**Fig. 8.** Conditional domain-adversarial network.

- **Physics-Informed Adversarial Network (PIAN)** is a DA framework that integrates physical knowledge into adversarial learning to improve transferability and interpretability in tasks where data are governed by underlying physical laws, such as structural dynamics or vibration-based SHM. The core idea is to combine a conventional adversarial DA architecture with additional physics-informed constraints that guide the learning process toward physically consistent representations. The adversarial component ensures domain-invariant feature alignment between source and target data, while the physics-informed component penalizes deviations from known physical relationships, such as governing equations, mode-shape consistency, frequency ratios, or energy conservation. These physics-based losses can be implemented in various forms, including explicit residual terms from partial differential equations, modal properties derived from structural models, or regularization based on physical parameter bounds. By embedding physical priors into the network, PIAN reduces reliance on purely data-driven statistical alignment and prevents the model from learning spurious correlations that violate physical realism. This hybrid learning strategy enables more robust cross-domain transfer, especially in scenarios with limited labeled data or strong distribution shifts caused by structural or environmental variations. Compared with standard adversarial networks, PIAN improves both interpretability and generalization in physics-governed systems, though it requires domain-specific knowledge and careful formulation of physics-based constraints to ensure stable joint optimization. [125].

Overall, these models illustrate the evolution of DA, from early statistical projection methods to deep adversarial and physics-guided frameworks. Each contributes unique strengths in addressing different types of domain shifts, supporting the long-term goal of achieving transferable and interpretable SHM under diverse operational conditions.

# 3. Applications of domain adaptation in SHM

Domain adaptation has been increasingly applied to address various distribution shifts that arise in practical SHM scenarios. Differences between training and deployment data often result from environmental variations, structural discrepancies, sensor configuration changes, or mismatches between simulated and real responses. These discrepancies can significantly degrade the performance of conventional machine-learning models trained under fixed conditions.

DA provides a unified framework to mitigate these challenges by transferring diagnostic knowledge across domains with differing characteristics. Through appropriate alignment or adaptation, models trained under one condition can remain effective when applied to another. The following subsections present representative applications of DA across vibration-based, ultrasonic, and hybrid monitoring contexts, highlighting how these methods enhance robustness and generalization in real-world SHM.

## 3.1 Environmental variations

Environmental variations, including temperature and seasonal changes, operational variability, and measurement inconsistencies, are major sources of domain shift in SHM, as shown in Fig. 9. These factors continuously modify signal distributions and sensor behavior even when the structure remains undamaged, often masking early signs of deterioration. Recent advances in DA have shown that learning domain-invariant representations can effectively compensate for such variability. By aligning features across environmental states and suppressing non-damage-related drift, these approaches enable SHM models to maintain reliability and discriminative power under evolving temperature, loading, and sensing conditions.

**a) Temperature effects**

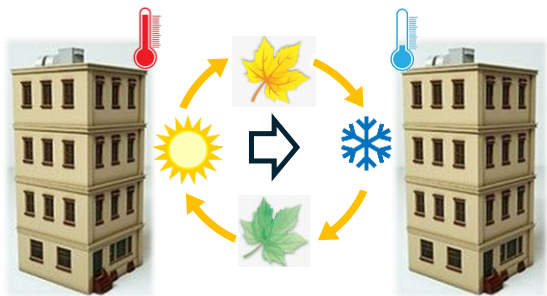

➤ Sensor damage

➤ Material Property Changes

➤ Thermal expansion and contraction

**b) Operational variability**

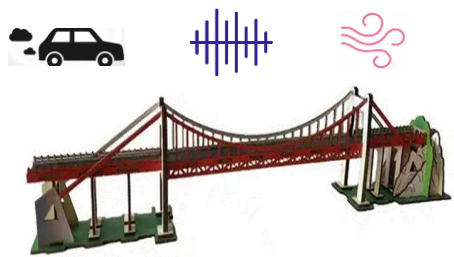

➤ Traffic load variation

➤ Machinery-induced variability

➤ Wind-induced variability

➤ Pedestrian-induced variability

**Fig. 9.** DA applications in environmental variations.

### 3.1.1 Temperature and seasonal effects

Temperature variations are among the most challenging and recurrent sources of domain shift in vibration-based SHM. Changes in temperature influence the material stiffness, mass density, damping ratio, and boundary conditions of structural systems, which in turn cause observable drifts in modal frequencies, mode shapes, and impedance responses. As a result, the same structure can exhibit significantly different feature distributions under varying temperatures even when undamaged. This thermal sensitivity severely degrades the performance of supervised learning models trained at fixed reference temperatures, leading to false alarms or missed detections. To address this, recent studies have adopted DA–based TL frameworks that align data collected under different temperature states, enabling damage recognition to remain robust across seasons and environmental variations.

Sawant et al. [85] presented an unsupervised transfer learning (TL) framework for temperature-compensated damage identification and localization using ultrasonic guided waves. Their approach first extracted damage-sensitive statistical features from guided-wave signals measured at multiple temperatures and then implemented a DA module to align the distributions of healthy and damaged states between baseline and target temperature domains. By minimizing the discrepancy between the source (reference temperature) and target (arbitrary temperature) domains, the framework successfully localized simulated damage without any labeled data at unseen temperatures. Experiments on a PZT-instrumented aluminum plate verified that the unsupervised DA model maintained high localization accuracy over a thermal range of 10–50 °C, proving the feasibility of learning domain-invariant representations for guided-wave signals affected by temperature drift.

A related impedance-based approach was proposed by Silva et al. [106], who introduced a TCA-based method to compensate for temperature effects in electro-mechanical impedance (EMI) measurements. The authors computed the root-mean-square deviation (RMSD) indices of both real and imaginary impedance components as discriminative features and considered them as samples from a labeled source domain and unlabeled target domains. Through TCA projection, these features were mapped into a latent subspace where thermal influence was minimized while structural state separability was preserved. A Mahalanobis distance-based classifier trained solely at the reference condition was then able to identify damage states under new temperatures without retraining. Laboratory validation on a PZT-bonded aluminum specimen exposed to controlled heating confirmed that the proposed framework efficiently transferred diagnostic knowledge between temperature domains, distinguishing actual structural degradation from temperature-induced impedance shifts.

Beyond deterministic subspace projection, probabilistic models have also been introduced to handle temperature-driven variability. Gardner et al. [97] proposed the Domain-Adapted Gaussian Mixture Model (DA-GMM) for population-based SHM, which learns a linear transformation aligning unlabeled target data with a Gaussian mixture model fitted to labeled source data. The adaptation is optimized via expectation–maximization to maximize the likelihood consistency between domains. Evaluations using the Z24 and KW51 bridge datasets showed that DA-GMM could effectively harmonize feature distributions across varying thermal and operational conditions, allowing accurate health-state inference without labeled target samples. This probabilistic alignment framework offers a natural treatment for uncertainty and noise, which are unavoidable under real environmental fluctuations.

A more computationally efficient statistical alignment strategy was proposed by Poole et al. [98], who formulated the Statistic Alignment method to equalize the mean and covariance of features between source and target domains. The authors developed two variants to progressively capture first- and second-order statistical relationships between the two domains, which are Normal-Condition Alignment (NCA) and Normal-Correlation Alignment (NCORAL). Their study demonstrated that NCORAL effectively reduced environmental and temperature-induced drifts in both numerical simulations and field bridge datasets, offering competitive accuracy with substantially lower computational demand compared with kernel-based or adversarial methods. This finding highlights that even simple statistical realignments can substantially enhance cross-temperature consistency in SHM data.

Temperature compensation is also crucial for heritage structures where seasonal cycles lead to modal frequency fluctuations that obscure damage patterns. Cavanni et al. [74] proposed a DA methodology for continuously monitored historical domes, employing manifold-based feature transformation to distinguish reversible thermal variations from irreversible stiffness losses. The method was applied to long-term monitoring data from masonry domes subjected to yearly temperature oscillations, successfully isolating temperature-related modal drifts and improving the interpretability of health indicators. This study exemplifies how DA concepts can be extended beyond laboratory-scale experiments to real, massive civil structures with complex environmental responses.

In addition, Souza et al. [75] conducted a comprehensive study on unsupervised TL for bridge SHM using JDA, explicitly considering temperature and seasonal effects as a form of environmental domain shift. Their framework aligns both the marginal and conditional distributions between source and target bridges by iteratively refining pseudo-labels in the target domain. Testing on several prestressed concrete bridges which experiences a wide temperature range from 10 °C to 36 °C showed that JDA effectively mitigated temperature-induced misclassification. The adaptation increased binary damage classification accuracy from below 30 % in the raw feature space to above 90 % after alignment, highlighting its robustness for cross-seasonal transfer in real monitoring campaigns.

Taken together, these studies demonstrate that temperature and seasonal variations can be effectively addressed through DA strategies that realign data across thermal conditions. Statistical approaches such as TCA, JDA, SA, and DA-GMM consistently reduce temperature bias by minimizing distributional discrepancies while preserving discriminative structural features. Their successful application across impedance, guided-wave, and vibration modalities confirms that DA has become a foundational technique for environmental compensation in SHM, ensuring that diagnostic models trained under controlled conditions remain reliable when deployed in real, temperature-varying field environments.

### 3.1.2 Operational variability

Operational variability, encompassing fluctuations in traffic loads, vehicle speed, boundary conditions, and excitation levels, is one of the principal factors that induce non-stationarity in structural responses and hence lead to domain shifts in data-driven SHM. Unlike temperature, which follows seasonal and predictable patterns, operational variability arises dynamically from stochastic excitations such as passing vehicles, train–track interactions, or wind-induced vibrations. These fluctuations alter the energy content and frequency characteristics of one-dimensional vibration signals, challenging the reliability of machine learning (ML) models trained under specific load or speed regimes. DA techniques have therefore been introduced to enable consistent performance by aligning feature distributions across distinct operational domains, thus ensuring robust damage diagnosis under varying service conditions.

Rezazadeh et al. [60] provided a broad systematic review of SHM methodologies under environmental and operational variability, highlighting DA and TL as effective strategies to achieve baseline independence. Their analysis classified mitigation approaches into three main categories: direct baseline compensation, adaptive/multi-baseline strategies, and reference-free DA techniques. Within the latter, unsupervised and semi-supervised DA frameworks were emphasized for their ability to accommodate load- or speed-induced response variations without requiring retraining for each operational state. The review synthesized how models trained on limited baseline data can generalize to unseen operational regimes through subspace-based and probabilistic alignment, confirming DA as a key enabler for operationally resilient SHM systems.

To address the specific challenge of domain shifts in bridge dynamics induced by variable loading and excitation patterns, Ardani et al. [86] proposed a transfer learning framework combining Proper Orthogonal Decomposition (POD) with DA. The study focused on vibration responses from an operational highway bridge subjected to varying traffic and environmental conditions. By extracting dominant modal components through POD and subsequently adapting the subspace representation between source and target datasets, the method preserved structural modes while filtering out operationally induced variability. Results demonstrated that even with substantial differences in vehicle-induced vibration amplitudes and traffic flow rates, the proposed POD–DA integration maintained consistent classification performance for structural condition states, outperforming conventional static baseline methods.

In a parallel line of research, Ghiasi et al. [61] introduced an unsupervised DA framework for drive-by monitoring of multiple railway tracks. Recognizing that vibration responses collected from in-service trains differ significantly between track segments due to operational factors such as speed, axle load, and rail stiffness, the authors implemented a three-stage pipeline consisting of feature extraction, DA, and damage diagnosis. The framework utilized acceleration data measured from bogie and car body sensors on a high-speed diagnostic train traversing multiple French railway lines. Four DA algorithms—Information-Theoretical Learning, TCA, GFK, and SA—were benchmarked. The study showed that unsupervised DA increased anomaly detection accuracy by 14% compared with non-adaptive unsupervised learning. The robustness of the framework was further validated under varying sensor layouts and partial label scenarios, demonstrating its ability to generalize across operational domains without additional training data.

Expanding on drive-by concepts, Ghiasi et al. [62] developed a progressive distribution alignment approach for railway condition monitoring using label correction–based adaptation. The proposed method introduced a stepwise correction mechanism that iteratively refined pseudo-labels in the target domain, improving alignment between source and target data representing distinct train operation speeds and dynamic loads. Acceleration signals collected from in-service trains were processed through multi-level DA layers, progressively minimizing discrepancies in both marginal and conditional distributions. Compared with standard subspace DA methods, this progressive strategy achieved more stable convergence and higher classification accuracy, particularly when the operational differences between source and target rail lines were pronounced.

Operational variability has also been investigated within the context of model-based SHM for complex truss systems. Huang et al. [76] proposed a digital-twin–enabled DA framework for truss bridge damage identification. The digital twin generated simulated vibration responses under different loading patterns, which were then aligned with field data through a feature transfer network based on domain-invariant representation learning. This alignment allowed the digital model, originally calibrated under nominal operation, to remain valid for diagnostic inference under variable operational conditions such as changing traffic loads or boundary interactions. The hybrid use of simulated and measured data proved particularly effective for systems where obtaining labeled data across all operational regimes is impractical.

Collectively, these studies demonstrate that DA offers a powerful strategy to mitigate the adverse effects of operational variability in SHM. Whether through subspace projection, statistical distribution matching, or digital twin coupling, DA consistently improves generalization under fluctuating load and speed conditions. The integration of unsupervised and semi-supervised paradigms enables the transfer of knowledge across operational domains without extensive labeled datasets, making these methods scalable for large infrastructure networks such as bridges and railways. Importantly, all these methods operate on one-dimensional vibration signals, demonstrating that the core DA principles originally developed for image analysis can be effectively reformulated for dynamic structural response data.

## 3.2 Sensor configuration

Sensor arrangement strongly influences the quality, completeness, and stability of data collected in SHM. Even when environmental and operational conditions are consistent, differences in sensor configuration can create substantial shifts in data distributions that limit model transferability. DA offers a means to mitigate these effects by aligning information across varying sensing layouts and data densities.

Visualized in Fig. 10, the challenge arises in two main forms. One concerns the spatial layout of sensors, where changes in position, spacing, or number affect how structural responses are captured. The other involves data sparsity and missing channels caused by sensor failure, data loss, or limited instrumentation. Recent studies show that by incorporating DA into both layout optimization and data reconstruction, models can retain consistency and sensitivity to damage under changing or incomplete sensor networks.

**a) Sensor layout differences**   **b) Data sparsity and missing channels**

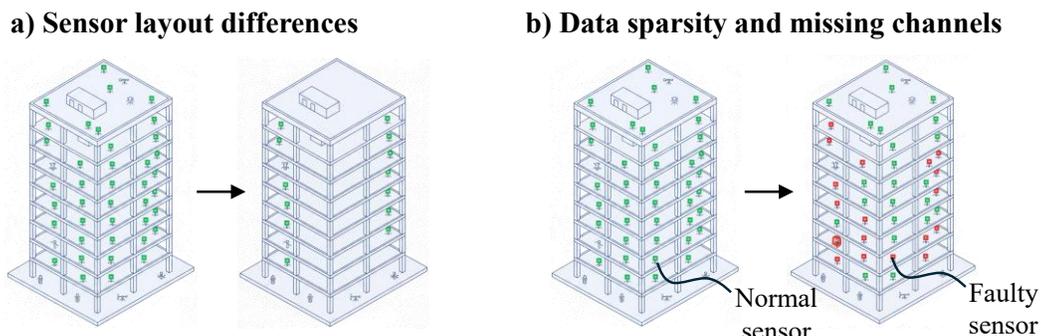

Normal sensor   Faulty sensor

**Fig. 10.** DA applications in sensor configuration.

### 3.2.1 Sensor layout

Sensor layout plays a critical role in the performance of vibration- and wave-based SHM systems. The spatial configuration, density, and placement of sensors directly determine the quality and diversity of measured responses, which are essential for accurate feature extraction and damage localization. However, in practical monitoring scenarios, sensor deployment is often constrained by accessibility, cost, and structural geometry, leading to limited or non-optimal layouts. Such variability can introduce domain discrepancies in the collected signals, as datasets recorded under different sensor configurations no longer share identical feature distributions. Consequently, DA techniques have been increasingly adopted to transfer diagnostic knowledge between different sensor networks, enabling models trained with one layout to generalize effectively to others.

Pan et al. [87] addressed sensor-related inconsistencies in monitoring data through a transfer learning-based anomaly detection framework. Their method combined CNNs with transfer learning to identify sensor-related data faults across different bridge structures. A CNN model was first trained using multivariate SHM data from one long-span bridge, including acceleration, strain, displacement, humidity, and temperature. The trained model parameters were then transferred to a second bridge using only a small subset of its data for fine-tuning. By freezing early convolutional layers and retraining only the final classification layers, the method successfully captured the shared anomaly features induced by similar sensor faults despite differing layouts. This transfer strategy significantly reduced the need for large labeled datasets and proved particularly useful in networks where sensor placement differed across spans or structural components, achieving higher fault-detection accuracy and improved generalization.

To overcome the limitation of data scarcity caused by sparse or uneven sensor placement, Chamangard et al. [99] proposed a compact 1D CNN-based transfer learning framework for damage detection in civil structures. Their approach used acceleration signals collected from a numerical model, an experimental structure, and a full-scale bridge, simulating progressively different sensor configurations. A compact CNN trained on one structure acted as the source network, and its learned weights were transferred to another structure with a different number and arrangement of sensors. Fine-tuning only the higher layers allowed the target model to achieve nearly 95% accuracy in damage detection even with limited data. The study demonstrated that transfer learning can effectively bridge domain gaps caused by inconsistent sensor distributions, ensuring robustness of vibration-based damage detection models across sensor-limited monitoring systems.

Building on the challenge of layout heterogeneity, Pinello et al. [63] proposed a Multilinear Principal Component Analysis (MPCA) -based DA framework tailored for ultrasonic guided wave monitoring. Their study explicitly addressed differences in both sensor type and array geometry by constructing a unified tensor representation combining the source and target sensor networks. MPCA was applied jointly to extract shared latent features that preserved spatial relationships among sensors while reducing the influence of layout differences. The adapted feature representations were then fine-tuned through a CNN regression model for damage localization. Testing on multiple composite plates with circular and rectangular sensor arrays demonstrated that the MPCA-based adaptation successfully transferred knowledge between sensor networks without manual dimensionality tuning. This approach achieved a substantial reduction in localization error compared to standard transfer learning techniques and effectively handled cases where the number or spatial distribution of sensors varied between domains.

Beyond laboratory-scale sensor arrays, Wan et al. [64] developed a DA-based identification method for vortex-induced vibrations (VIV) of long-span bridges, where sensor configurations were inherently uneven due to span geometry and accessibility. The authors integrated unsupervised DA into a DL framework to align feature distributions across different sensor groups monitoring the bridge deck and main cables. By exploiting unlabeled acceleration and strain data from multiple locations, the proposed approach achieved accurate VIV identification without prior assumptions about the sensor network layout. This strategy proved particularly advantageous for large infrastructures, where practical monitoring constraints often lead to asymmetric or incomplete sensor coverage, highlighting the importance of DA in harmonizing heterogeneous measurement configurations.

Finally, Sajedi et al. [100] introduced a deep generative Bayesian optimization framework for optimal sensor placement in SHM. Although not a DA method in the strictest sense, this approach is closely related to sensor layout optimization through its data-driven integration of generative modeling and Bayesian inference. The framework employed deep generative models to predict the expected information gain of potential sensor placements and used Bayesian optimization to identify configurations that maximized diagnostic performance while minimizing sensor redundancy. By learning from existing sensor configurations and simulated vibration data, the model adapted to new structures and provided layout recommendations that balanced sensitivity and cost-efficiency. This generative optimization method complements DA-based strategies by proactively designing sensor layouts that minimize future domain discrepancies across monitoring systems.

In summary, DA and transfer learning techniques provide a powerful foundation for overcoming the challenges posed by varying sensor configurations in SHM. Methods such as CNN fine-tuning, MPCA-based feature alignment, and unsupervised adaptation have demonstrated strong capability to transfer diagnostic knowledge across heterogeneous sensor networks. These approaches allow for the unification of feature spaces between different sensor layouts, improving consistency and reliability in vibration- and wave-based monitoring. As demonstrated across bridges, plates, and composite structures, DA not only compensates for incomplete or inconsistent sensor distributions but also enables scalable SHM solutions where retraining for every layout becomes unnecessary.

### 3.2.2 Data sparsity and missing channels

Incomplete or uneven sensor measurements often arise in long-term monitoring due to sensor malfunction, communication loss, or spatial sparsity. Such missing channels can distort the underlying data distribution, making it difficult for supervised models to maintain accuracy when applied to partially observed signals. Recent research has applied DA and transfer learning to recover or realign incomplete vibration or guided-wave datasets, allowing models trained on complete data to generalize effectively to degraded measurement conditions.

Liao et al. [116] proposed a Discriminative Wavelet Adversarial Adaptive Network (DWAAN) for cross-scene guided-wave debonding detection in reinforced concrete beams. Their approach tackled the problem of varying sensor configurations and incomplete wave paths across different structures. The framework employed continuous wavelet transforms to convert one-dimensional ultrasonic responses into time–frequency features, followed by adversarial and statistical alignment between source and target domains. A domain discriminator and MMD loss jointly enforced distribution consistency, enabling the model trained on one beam to detect interfacial debonding in another. Experimental and numerical validations showed that the adapted network improved detection accuracy by over 25% compared with baseline CNNs, demonstrating that DA can mitigate the effects of missing or shifted sensor locations in guided-wave monitoring.

Zheng et al. [65] addressed the issue of incomplete vibration data in bridge monitoring by developing an unsupervised DA–based imputation framework. The method first pretrained a reconstruction network on synthetically masked vibration sequences and then fine-tuned it through adversarial alignment between the synthetic and real incomplete domains. The architecture used discrete wavelet decomposition to capture multi-scale temporal information from one-dimensional acceleration signals. By transferring the learned mapping to real missing data without explicit mask information, the model successfully reduced the average reconstruction error by nearly one-third compared with conventional GAN-based imputers. The study demonstrated that cross-domain feature alignment can effectively restore information from partially missing sensor channels, preserving the structural response patterns critical for damage inference.

In summary, these two studies show that DA can be extended beyond damage classification to handle sensor sparsity and missing-channel issues. Whether through adversarial feature alignment in guided-wave testing or through generative imputation for incomplete vibration sequences, both approaches demonstrate that learning transferable representations across measurement domains enhances the robustness of SHM systems under realistic, imperfect sensing conditions.

## 3.3 Structural differences

Beyond environmental and operational variations, intrinsic structural differences pose a more fundamental source of domain shift in SHM. Even for nominally similar assets variations in geometry, stiffness, boundary conditions, and material composition can induce significant discrepancies in dynamic response. These discrepancies distort modal frequencies, transmissibility functions, and spatial vibration patterns, thereby breaking the statistical assumptions required for supervised models trained on a single configuration. DA offers a principled way to address such challenges by learning representations that are invariant to structural configurations while preserving sensitivity to damage-relevant features. Recent advances in deep and probabilistic adaptation have enabled knowledge transfer not only between individual pairs of structures but also across entire populations. The former, often termed one-to-one transfer, focuses on aligning features between two physically related structures, whereas the latter, population-to-one transfer, aggregates multiple heterogeneous sources to form generalized latent spaces capable of adapting to unseen targets. These challenges can be identified in Fig. 11. Together, these developments mark a shift from case-specific calibration toward structure-agnostic SHM models that learn universal, physically consistent representations across diverse infrastructures.

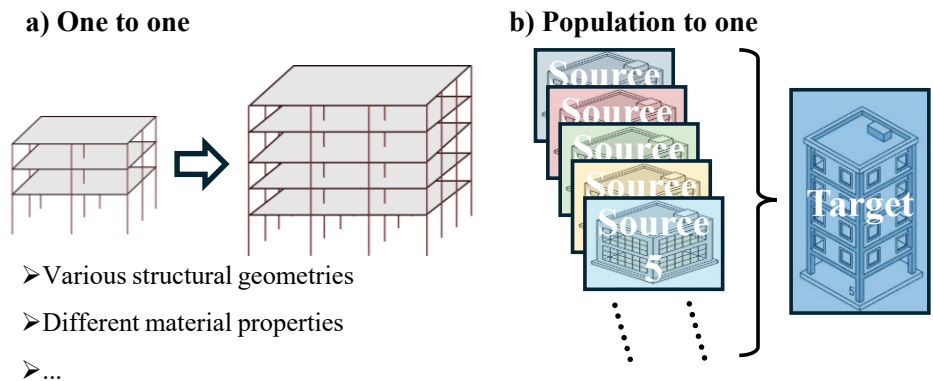

**a) One to one**

➢ Various structural geometries
➢ Different material properties
➢ ...

**b) Population to one**

**Fig. 11.** DA applications in structural differences.

### 3.3.1 One-to-one transfer

In one-to-one transfer for bridge SHM, knowledge is moved from a single source bridge to a single target bridge that differs in geometry, boundary conditions, or sensing configuration. The core requirement is to transform vibration-based features into a shared representation where the inter-bridge distribution shift is reduced but damage sensitivity is preserved. Two complementary routes are common: (i) feature-based DA that explicitly aligns source–target statistics, and (ii) deep transfer learning that couples 1D/2D convolutional feature extractors with alignment losses or fine-tuning, sometimes using simulation-to-experiment pairings to compensate for scarce labels on the target.

To anchor one-to-one transfer in bridges, it is useful to first clarify when transfer is feasible and what feature–alignment choices matter in practice. Yano et al. [40] laid out foundations and applicability of transfer learning for bridges and benchmarked three feature-based methods using real monitoring data from Z-24 and PI-57/PK 075+317; treating one bridge as labeled source and another as unlabeled target, they showed that simple modal-frequency features aligned via JDA/MIDA/TCA can deliver effective cross-bridge classification, thereby formalizing when one-to-one transfer is feasible and how the choice of modal features and alignment influences performance.

With these foundations in place, a first concrete demonstration aligns labeled modal features from one bridge to an unlabeled counterpart to achieve cross-bridge damage classification. Giglioni et al. [77] proposed a DA approach that aligns damage-sensitive modal features collected over long-term monitoring between two bridges (Z24 and S101) and, in a parallel study, between their finite element models; the method minimizes the source–target distance in a shared feature space so a supervised classifier trained on the labeled source can generalize to the unlabeled target, demonstrating successful exchange of health-state labels for bridge damage classification under intrinsic structural differences. Their implementation explicitly uses natural frequencies as features and evaluates Joint Domain Adaptation and Normal Condition Alignment within a unified pipeline for cross-bridge transfer.

Beyond direct statistical alignment, an alternative is to translate structural states between dissimilar bridges while preserving condition information. Luleci et al. [117] introduced structural state translation (SST), a domain-generalization pathway that learns a condition-preserving translation between two prestressed concrete bridges without requiring label transfer; trained on two conditions (State-H/State-D) from Bridge #1 using a Domain-Generalized Cycle-Generative model, the system "translates" Bridge #2's states and validates them against modal identifiers and mean magnitude-squared coherence, achieving close agreement. Thus enabling condition transfer between dissimilar bridges in a one-to-one setting via learned state translation rather than direct statistical alignment.

Moving from hand-crafted modal features to learned representations, deep convolutional transfer can align both global and class-conditional distributions. Chen et al. [88] proposed a deep convolutional transfer framework that combines one-dimensional and two-dimensional CNN branches to mine spatiotemporal features from raw vibration data, and developed a DA module that couples multi-kernel MMD with local MMD to align both global and subdomain distributions between two structures; transfer experiments on two different test structures verified that aligning global and local distributions markedly improves target recognition when the target bridge lacks labels, illustrating a deep 1D/2D feature route to one-to-one alignment for vibration signals.

A closely related challenge is transferring from simulation to experiment, where time–frequency representations feed a fine-tuned deep architecture. Lu et al. [66] addressed one-to-one transfer between a simulated cable-stayed bridge and an experimental reduced-scale specimen by proposing a DA model, SE-TL-ResNet34; acceleration time series are converted to time–frequency images via S-transform, a CNN extracts modal damage features, and a pre-trained network is fine-tuned to classify multiple damage types on the target, outperforming traditional supervised baselines and showing that simulation-to-experiment alignment with attention-enhanced residual blocks can deliver accurate damage identification under noise and limited target data.

In longitudinal monitoring, one-to-one transfer must also accommodate structural modifications such as retrofitting without relabeling. Yano et al. [89] presented a study on transfer learning for SHM in bridges that underwent retrofitting, using real monitoring data from two European railway bridges. The work addressed the practical challenge of reusing classifiers trained on pre-retrofit data for post-retrofit conditions, where structural stiffening alters modal frequencies. By applying a JDA method, the authors aligned the statistical distributions of vibration features before and after retrofitting, allowing the same outlier-detection classifier to remain effective. Case studies on the PK075+317 bridge in France and the KW51 bridge in Belgium demonstrated that transfer learning enables long-term SHM continuity across structural modifications without relabeling, validating its capability to manage temporal domain shifts within one-to-one transfer contexts.

When both source and target are real bridges, feature-based alignment remains competitive for pedestrian spans under operational variability. Marasco et al. [78] developed an unsupervised transfer learning framework for pedestrian bridges, exploring how knowledge from one or a small group of bridges can generalize to another with similar geometry. Using both numerical models and experimental vibration data from two prestressed concrete footbridges in Lisbon, they compared z-score normalization and TCA for aligning features across bridges. The TCA-based mapping outperformed normalization by minimizing statistical discrepancies between domains while retaining physical interpretability of modal features. The study confirmed that transfer learning can extend SHM to new, unmonitored bridges—an experimental-to-experimental one-to-one transfer that reduces the need for independent training on every structure.

Drive-by scenarios further motivate hierarchical feature disentanglement so that span-level and bridge-level patterns can be adapted jointly. Liu et al. [118] proposed Hierarchical Multi-task Unsupervised Domain Adaptation (HierMUD), a drive-by framework that transfers knowledge across bridges in a hierarchical feature space. Each layer learns bridge-specific and global features from vehicle-induced acceleration responses, while an adversarial loss ensures alignment between the source and target bridge domains. The hierarchical design captures both local span-level and global bridge-level representations, which are then adapted to a single target bridge for damage diagnosis. Results indicated substantial gains in transfer accuracy and stability compared with shallow or single-level adaptation, confirming that multi-level feature disentanglement enhances one-to-one bridge transfer under dynamic excitation.

Even within a single instrumented bridge, domain adaptation principles enable transfer across spans with different boundary conditions. Giglioni et al. [79] reported a DA case study on bridge spans using experimental data from a steel–concrete composite bridge. Their approach combined measured vibration data and finite-element predictions to perform one-to-one knowledge transfer across adjacent spans. Using correlation alignment between modal features, the framework achieved consistent damage classification accuracy across spans despite differences in geometry and boundary conditions. The experiment highlighted that even within a single bridge, domain adaptation principles can transfer knowledge between substructures, exemplifying the granularity achievable within one-to-one transfer methodologies.

Finally, probabilistic transfer and calibration can complement alignment by accounting explicitly for model and measurement uncertainty. Ferreira et al. [67] introduced a Bayesian transfer learning and calibration framework to address uncertainty and limited labeled data for twin concrete bridges. In their setup, one bridge served as a well-instrumented source, and its probabilistic model parameters were transferred to a second, similar target bridge using Bayesian updating. Acceleration-based modal frequencies and damage indices served as the shared features for domain bridging. The method reduced data demands while maintaining predictive accuracy, showing that probabilistic transfer can complement feature-based adaptation by accounting for uncertainty in model and measurement noise during cross-bridge transfer.

Integrating these five works with the earlier studies, one-to-one transfer in bridge SHM now spans both feature-level alignment and deep-learning-based adaptation. Collectively, they show that knowledge can be reliably migrated between a pair of bridges or between different states, through systematic distribution alignment of vibration-derived features. The success of these studies underscores that even when bridges differ in geometry, materials, or temporal conditions, DA allows vibration-based models trained on one structure to remain valid on another. This one-to-one transfer paradigm forms the methodological foundation for scaling SHM toward broader population-level applications.

### 3.3.2 Population-to-one transfer

Population-to-one transfer in SHM aims to generalize knowledge learned from a population of bridges to a single target bridge that has few or no labeled data. Instead of treating each structure independently, these methods regard all bridges as members of a collective system sharing similar physical principles but differing in geometry, materials, or boundary conditions. By learning domain-invariant representations from multiple sources, the diagnostic model becomes capable of identifying damage on a new target bridge even when its dynamic responses exhibit distributional discrepancies. The central task is to reduce the feature mismatch between multiple sources and the target while maintaining sensitivity to damage-related patterns derived from one-dimensional vibration signals.

For population-to-one transfer, it is essential to clarify the theoretical conditions under which knowledge can be pooled from many sources and applied to a single target. Gardner et al. [109] provided a theoretical foundation for applying DA to population-based SHM. They formulated the transfer learning process as a mapping of vibration-based feature distributions from multiple labeled sources to an unlabeled target. Their analysis clarified that transfer is appropriate when the feature and label spaces are compatible but the marginal or conditional distributions differ. Using vibration and modal data from several case studies, they demonstrated that transfer learning can preserve the physical meaning of features while minimizing the statistical distance between domains. This work established a general framework explaining why multi-source training stabilizes the latent representation used for damage detection on a single target bridge.

Building on this rationale, a laboratory study demonstrates that training on multiple spans yields more stable target performance than any single-span source. Giglioni et al. [68] performed one of the first experimental demonstrations of population-based transfer learning for bridges. Using a laboratory setup of multi-span continuous girder bridges, they investigated how classifiers trained on vibration features from several spans could be adapted to an unseen span through domain alignment. The study evaluated different combinations of source and target domains, showing that models trained on multiple spans were more robust than those derived from any single configuration. The bridge acceleration responses were processed into modal-frequency features and used for damage classification across the population. Their findings confirmed that DA allows information from a population of similar bridges to be transferred effectively to one target structure, establishing the feasibility of multi-source-to-one transfer in practical monitoring environments.

Extending feasibility from similar bridges to heterogeneous populations requires an explicit mapping framework across geometries and topologies. Gardner et al. [107] developed heterogeneous population-based DA, which laid the mathematical groundwork for mapping and knowledge transfer across structurally diverse systems. They formalized DA as a mapping between source and target feature spaces within a population, using topology-based graph representations of structures. The paper defined when DA is feasible by examining geometric, material, and topological differences and showed that vibration-derived modal features could be transferred between members of a heterogeneous population via graph-based correspondence. This theoretical framework underpins later empirical studies of bridge SHM, providing criteria for when multi-bridge transfer is valid and stable.

Beyond alignment, condition translation enables population models to carry health states across structurally distinct members. Luleci et al. [123] introduced a structural state translation framework that learns how to map structural conditions from one bridge to another without the need for target labels. Their method, referred to as Structural State Translation, relies on a domain-generalization model that captures invariant relationships between vibration responses measured under different conditions. Once trained, the model can translate the condition of a target bridge into the latent representation of a source bridge, enabling cross-structure health assessment. Experimental results on prestressed concrete bridges showed that the translated states preserved modal frequencies and mode shapes, demonstrating that condition transfer between structurally distinct bridges is possible through learned state representations derived from acceleration signals.

Domain generalization further separates structure-specific factors from damage-related cues when the target has no data. Nie et al. [69] developed a deep domain-generalization network to improve structural damage detection when data come from multiple bridges with diverse properties. Their approach separates structure-specific variations, such as material or geometric differences, from damage-related features using a combination of normalization and feature disentanglement techniques. The network was trained on acceleration-based time–frequency representations collected from several bridges and then applied to a new bridge that was not included in the training data. The results showed high cross-domain accuracy, indicating that the model learned generalized features that remain stable under new structural conditions. This study confirmed that domain generalization provides a viable way to perform population-to-one transfer when direct target data are unavailable.

When multiple sensors are available, multi-channel adaptation exploits complementary dynamics while aligning sources and the target. Xiao et al. [101] proposed a multi-channel domain-adaptation deep transfer learning framework to perform damage diagnosis across bridges with unlabeled target data. Their model uses vibration signals collected from multiple sensors as one-dimensional time series, which are processed through convolutional layers to extract spatial–temporal features. To ensure that knowledge from multiple bridges could be transferred effectively, a DA component based on multi-kernel MMD was used to align the feature distributions of all sources and the target. The model was trained using data from three bridges and successfully identified damage on the target bridge. This work highlighted that combining multichannel vibration data with statistical alignment techniques can realize robust population-to-one transfer in practical bridge monitoring scenarios.

Sub-domain alignment then refines class-conditional matching to handle local distribution shifts across bridges. Xiao et al. [102] introduced a sub-domain adaptive deep transfer learning network (SADTLN) to handle distribution discrepancies between bridges with different geometries and loading environments. The model employs a multi-kernel local maximum mean discrepancy (MK-LMMD) loss combined with a domain classifier to align both global and sub-domain feature distributions. Vibration-based signals collected from multiple bridges were used as one-dimensional inputs to a CNN extractor, ensuring that the learned features were domain-invariant while retaining category discrimination. The method effectively recognized unlabeled health states on new bridges, demonstrating robust population-to-one transfer under realistic variations.

In parallel, simple statistical normalization of normal conditions can already enable robust transfer when damage-sensitive parameters correspond physically. Giglioni et al. [80] presented population-based DA method, which validated a statistical-alignment technique named NCA for transferring knowledge between dissimilar bridge models. Using two finite-element bridges (Z24 and S101) representing different materials and geometries, the study aligned the natural-frequency features of healthy and damaged states across domains. By matching lower-order statistics of the feature distributions, the classifier trained on the source bridge achieved consistent performance on the target. This work demonstrated that simple linear alignment in the original feature space can be sufficient when the damage-sensitive parameters share physical correspondence, highlighting an interpretable, data-efficient path to population-level transfer.

Joint training across simulated sources stabilizes deep features and mitigates forgetting when facing unseen target configurations. Duran et al. [81] conducted leveraging DL for structural damage detection, focusing on integrating finite-element simulations with deep transfer learning. Acceleration time-history signals from multiple bridge-type structures were transformed into two-dimensional representations and fed to a CNN for damage classification. After observing accuracy degradation when testing on bridges with different geometries, they applied feature-extraction-based transfer learning and joint training across several source models. The joint-training strategy preserved generalization and prevented catastrophic forgetting, enabling reliable prediction of unseen damage scenarios. Although the study used grayscale-encoded signals, its essence lies in the transfer of vibration-based knowledge from simulated to heterogeneous structural domains.

Adversarial weighting helps prioritize reliable sources and down-weight uncertain samples within a heterogeneous population. Xiao et al. [124] proposed an adversarial sub-DA strategy that integrates fuzzy weighting into deep transfer learning. Their model employed adversarial training between a feature extractor and a domain discriminator to minimize the distance between multiple bridges' distributions while adaptively assigning fuzzy weights to uncertain samples from unlabeled target data. Bridge vibration signals served as one-dimensional time-series inputs. The fuzzy weighting mechanism allowed the model to focus on more reliable source samples during adaptation, yielding superior performance in identifying multiple unseen damage patterns. The study emphasized the importance of integrating probabilistic weighting and adversarial alignment for handling heterogeneous bridge populations with imbalanced or partially noisy data.

A multi-channel sub-domain strategy can further disentangle sensor-specific variations from shared damage signatures. Chen et al. [145] proposed multi-channel sub-DA to enhance cross-bridge knowledge transfer under significant structural differences. The model integrates a multi-channel 1D convolutional extractor with sub-domain adaptive modules that operate independently on channel-specific feature spaces. Each channel processes one-dimensional vibration signals collected from different sensor locations, capturing complementary structural dynamics. The method introduces a MK-LMMD loss to align conditional distributions at the sub-domain level, enabling the model to distinguish damage-sensitive features from domain-specific noise. By combining global and local adaptation, the framework achieved stable performance when transferring from a group of prestressed concrete bridges to an unseen target. The study emphasized that multi-channel sub-domain alignment offers an effective pathway for population-to-one transfer when vibration data exhibit high sensor heterogeneity and varying boundary conditions.

Injecting physics into multi-source adversarial adaptation improves interpretability and reduces negative transfer across dissimilar structures. Xu et al. [125] proposed a physics-informed multi-source domain adversarial network to achieve cross-building seismic damage diagnosis without requiring labeled data from the target structure. The framework integrates deep adversarial learning with physical knowledge of structural dynamics to enhance transferability and interpretability. By introducing a physics-guided weighting mechanism, the model assigns higher importance to source buildings with similar physical characteristics, such as stiffness and modal properties, thereby reducing the negative transfer caused by structural dissimilarity. A convolutional feature extractor and multiple domain discriminators are jointly trained through adversarial optimization to learn domain-invariant yet damage-sensitive features from one-dimensional vibration signals. Validation using both numerical simulations and experimental building data showed that the method significantly improved accuracy and stability compared with conventional DA approaches, highlighting the potential of combining physics-informed constraints and adversarial transfer for seismic damage identification.

Open-set extensions are needed to recognize novel target states while preserving alignment over shared classes. Xiao et al. [119] introduced adversarial auxiliary weighted subdomain adaptation for bridge damage diagnosis, addressing the challenge of unknown damage types appearing in the target domain. The proposed Adversarial Auxiliary Weighted Subdomain Network (AWSDN) integrates adversarial training with a multi-channel multi-kernel weighted local MMD (MCMK-WLMMD) metric to isolate target outliers while aligning shared sub-domains. One-dimensional bridge vibration signals serve as the model's direct inputs, while adversarial learning introduces auxiliary weights representing sample similarity between source and target distributions. This approach allows the system to identify previously unseen damage states and prevent negative transfer caused by outlier samples. Through experiments conducted on three real bridges in Japan, the framework achieved robust recognition of new damage scenarios, demonstrating its practical utility for population-to-one SHM under open-set conditions.

Finally, cross-domain pretraining from audio shows that cepstral features can enrich vibration representations when damage data are scarce. Tronci et al. [103] presents an unconventional approach that connects vibration-based SHM with techniques originally developed for speech recognition. The study addresses the scarcity of labeled damaged-state data in civil applications by transferring knowledge from a source domain with abundant audio data to a target domain composed of limited structural vibration records. Specifically, the framework first learns low-level feature representations from a large-scale speaker-recognition dataset (VoxCeleb), where x-vectors and Mel-Frequency Cepstral Coefficients (MFCCs) are extracted to capture discriminative spectral patterns. These learned representations are then fine-tuned using vibration signals from the Z24 bridge to identify damage states. By leveraging the shared statistical structure between audio and structural vibrations, the model gains a richer understanding of dynamic signal characteristics and demonstrates strong generalization across operational conditions. The results show that cepstral features, though originally designed for speech, can serve as effective damage-sensitive indicators, while transfer learning significantly mitigates data imbalance between healthy and damaged classes in bridge monitoring.

Collectively, these three works extend the frontier of population-to-one transfer learning in bridge SHM by addressing open-set adaptation, physical interpretability, and multi-channel heterogeneity. From sub-domain-wise alignment to adversarial weighting and physics-guided regularization, each framework refines how vibration-based knowledge is transferred between bridges. The results collectively confirm that embedding physical constraints, adaptive weighting, and multi-scale sub-domain learning enables reliable cross-bridge damage assessment, even when the target structure presents new damage types, unknown conditions, or unseen configurations.

## 3.4 Simulation-to-real transfer

Simulation-to-real transfer has become a critical frontier in SHM, bridging the gap between physics-based numerical models and data collected from real structures. While finite-element simulations can efficiently generate large, fully labeled datasets under controlled boundary and material conditions, these synthetic domains often diverge substantially from field measurements due to unmodeled damping, measurement noise, and environmental complexity. As a result, models trained purely on simulated data tend to overfit the idealized physics and fail when deployed in practice. DA provides a framework to reconcile this discrepancy by aligning the feature distributions or structural representations between simulated and real domains, ensuring that simulated knowledge remains useful for real-world inference. As shown in Fig. 12, recent works have integrated physics priors, generative adversarial learning, and self-supervised representation alignment to translate synthetic responses into realistic sensor-level signals while retaining damage semantics. Collectively, these approaches redefine simulation as an active, transferable source of knowledge rather than an isolated modeling tool—enabling scalable, data-efficient SHM pipelines where numerical simulations and field monitoring jointly contribute to robust, physically grounded damage assessment.

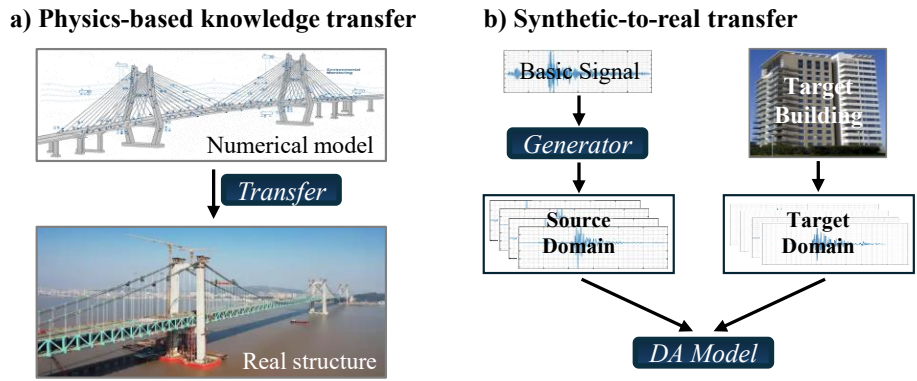

**Fig. 12.** DA applications in simulation-to-real transfer.

### 3.4.1 Physics-based knowledge transfer

Physics-based DA aims to bridge the gap between simulated and real vibration data by embedding structural mechanics into machine learning models. Instead of relying purely on data-driven mappings, these methods transfer physically meaningful features from finite element or analytically simulated structures to real bridges under operational conditions. This approach enhances interpretability and ensures that the learned representations obey known dynamic laws such as stiffness–frequency relationships and mode-shape consistency. Recent developments in this area demonstrate that the combination of DL and physics-based constraints enables accurate and generalizable structural diagnosis even when field data are sparse or contaminated with noise.

Talaei et al. [110] developed a hybrid approach that combines analytical bridge–vehicle interaction models with unsupervised DA to estimate prestressing forces in concrete bridges. The framework first simulates dynamic responses using a detailed physics-based model, generating labeled data under controlled loading scenarios. These are treated as the source domain, while field measurements recorded under unknown traffic excitations represent the target. A statistical alignment module minimizes the feature discrepancy between simulation and experiment, while a physical constraint term ensures consistency with stress–strain predictions from the prestressing model. Validation on an instrumented bridge demonstrated that the method could infer real prestress loss with high accuracy, illustrating how embedding structural equations within adaptation enhances physical interpretability.

Building on physics-embedded transfer, hybrid digital twins inject structural constraints directly into the adaptation process. Wang et al. [111] introduced a physics-informed machine learning framework that integrates digital twin concepts with transfer learning for structural damage detection and localization. A hybrid digital twin combining finite element simulations with real-time monitoring data is used to guide the network training process. Physical constraints derived from structural dynamics equations are embedded into the loss function to ensure that the learned features obey mechanical consistency while improving domain generalization. Through the integration of DA and physical regularization, the framework effectively localized damage in complex civil structures, outperforming purely data-driven models in both accuracy and interpretability.

To quantify damage under transfer, mapping with probabilistic regression provides continuous indices with uncertainty. Yano et al. [90] proposed a transfer learning method that combines TCA with Gaussian process regression to achieve damage quantification across different structural domains. The authors treated numerical simulation data from a benchmark three-story building as the labeled source and experimental vibration data as the unlabeled target. By mapping both domains into a common latent subspace, they reduced feature discrepancies while preserving label consistency, after which the Gaussian process regression predicted continuous damage indices with quantified uncertainty. This approach enabled a model trained on synthetic acceleration responses to successfully estimate real structural damage severity, confirming that probabilistic regression coupled with DA can handle nonlinear discrepancies between simulated and real dynamics.

Zhao et al. [104] combined transfer learning with Bayesian model updating to address modeling uncertainty in structural damage identification. The method begins by pretraining a neural network on data from a structurally similar source bridge to obtain informative parameter priors. Bayesian inference is then applied to update stiffness and damping parameters using new vibration responses from the target structure. This integration of deterministic learning and probabilistic updating enables the model to capture epistemic uncertainty while preserving adaptability across structural configurations. Numerical simulations and experimental bridge tests verified that this hybrid method yields faster convergence, improved parameter accuracy, and more reliable posterior estimates compared with conventional Bayesian updating alone. The study underscores how physics-guided probabilistic inference coupled with DA can produce transferable and trustworthy damage identification frameworks.

In practice, a data-efficient entry point is to pretrain on simulations and fine-tune with scarce field data. Bao et al. [91] proposed a deep transfer framework that utilizes simulated vibration signals as the source domain to improve structural condition identification when real-world data are limited. Their model integrates a convolutional feature extractor with a bidirectional LSTM network to capture both spatial and temporal dependencies in acceleration sequences. Pretraining is carried out using a large set of synthetic simulations, after which the model is fine-tuned on a small portion of real bridge data. This process allows the network to retain modal relationships learned from physics-based simulations while adapting to environmental variability present in field measurements. Testing on multiple bridge cases confirmed that pretraining on synthetic dynamics substantially enhanced fault recognition accuracy and reduced overfitting to noisy real data.

Lin et al. [105] introduced a dynamics-based deep transfer learning strategy to perform cross-domain damage detection between numerical and experimental structures. A convolutional neural network was designed to extract damage-sensitive and domain-invariant features from one-dimensional vibration signals. Unlike conventional methods that attempt to reduce the simulation-to-experiment gap by direct calibration, this approach trained a joint feature extractor using both domains simultaneously without requiring target labels. Numerical and laboratory experiments on beam structures demonstrated that this learning strategy effectively transferred knowledge from simulated to physical systems, yielding higher accuracy in identifying damage under environmental and modeling uncertainties.

Ghazimoghadam et al. [82] proposed an unsupervised DL strategy based on a multi-head self-attention LSTM autoencoder for vibration-based damage detection. The network learns damage-sensitive patterns from simulated acceleration responses and transfers them to experimental data through feature reconstruction. The self-attention layers assign higher weights to time segments associated with modal changes, thereby filtering out operational noise. DA is implicitly achieved by training the autoencoder to minimize the reconstruction error across both simulated and real datasets without explicit labels. Experiments using laboratory-scale bridges verified that this framework can generalize across different structural conditions while maintaining sensitivity to stiffness degradation.

Beyond reconstruction, conditional adversarial alignment jointly matches features and label structure across domains. Li et al. [113] proposed a conditional adversarial DA framework to perform cross-domain damage identification between simulated and experimental structures. The model employs a convolutional feature extractor to process one-dimensional vibration signals and an adversarial discriminator conditioned on class probabilities to achieve joint alignment of feature and label distributions. An entropy-based weighting scheme stabilizes training by suppressing uncertain samples that might introduce negative transfer. Evaluations using both numerical beam models and physical bridge experiments demonstrated that the method successfully transferred knowledge across domains with different stiffness and mass properties, accurately detecting damage even in the absence of target labels.

Ozdagli et al. [108] addressed structural fault detection under model uncertainty using a domain adversarial neural network. They generated synthetic labeled vibration signals from numerical simulations and used real unlabeled measurements as the target domain. The network consisted of a feature extractor, label predictor, and domain discriminator trained with gradient reversal to achieve domain confusion while retaining damage sensitivity. Two validation cases confirmed that this adversarial alignment enabled the model to accurately detect faults in experimental data, outperforming conventional transfer learning methods. The authors emphasized that embedding domain invariance directly into the learned features is key to bridging model–reality discrepancies in structural applications.

Partial domain adaptation further emphasizes shared classes while suppressing unrelated ones. Zhang et al. [70] developed a class-weighted subdomain adaptation framework to enhance the generalization of damage identification models trained on numerical data when applied to real structures. The model integrates class-level weighting into a convolutional neural network to account for the imbalance and partial overlap of label spaces between simulated and experimental datasets. A combined loss function composed of classification cross-entropy and local maximum mean discrepancy ensures that shared damage classes contribute more significantly to the alignment process, while unrelated ones are suppressed. The method achieves both subdomain alignment and partial DA, making it suitable for practical applications where real structural data cover only part of the simulated damage classes. Experiments on a cantilever beam under vibration excitation confirmed its superior robustness and higher damage identification accuracy compared with conventional DA schemes.

Zhang et al. [112] introduced an unsupervised DA approach based on transmissibility functions to perform vibration-based damage detection without requiring labeled measurements from real structures. Their network combines a feature generator, two classifiers, and a discriminator in a joint maximum classifier discrepancy and adversarial discriminative training process. Domain-level and class-level alignments are executed simultaneously to merge global and local feature distributions between numerical and experimental data. The transmissibility function is employed to mitigate the effects of unknown external excitations, ensuring that extracted features remain sensitive to structural damage while invariant to environmental noise. The proposed model was validated through three case studies involving buildings and the Canton Tower, demonstrating accurate identification of structural damage under unlabeled conditions.

Frequency-domain invariant representations help to narrow the simulation–field gap for track systems. Wang et al. [71] developed a frequency-domain invariant representation network to transfer knowledge from vehicle–track simulations to field tests for track damage identification. The framework used acceleration time histories from the vehicle–track coupled model as source data and real railway measurements as target data. Both datasets were transformed into the frequency domain, and an autoencoder-based feature extractor was trained to learn invariant latent representations that minimized domain discrepancy. Through this process, displacement was reconstructed from acceleration, enabling more reliable identification of stiffness degradation. The study demonstrated that frequency-domain invariant features yield stronger generalization between simulated and real responses, significantly improving diagnosis accuracy for track conditions.

Wang et al. [120] proposed an adversarial DA method based on finite element simulations for fatigue crack detection using Lamb waves. Simulated ultrasonic signals under multiple damage scenarios served as labeled source data, while experimental waveforms from fatigue tests formed the unlabeled target. The model combined maximum mean discrepancy with a domain-adversarial neural network to extract discriminative yet domain-invariant features, allowing classifiers trained on simulations to generalize to real measurements. Experiments on metal specimens confirmed that this hybrid adaptation scheme significantly improved cross-domain crack detection, establishing an effective simulation-to-real pipeline for Lamb wave-based SHM.

Simple TCA on FE–field features already stabilizes transfer under model mismatch. Figueiredo et al. [92] proposed an unsupervised transfer learning scheme for bridges that combines finite element simulations and field monitoring data to improve damage detection performance. In this framework, classifiers are trained solely on simulated vibration features generated by an FE model and are subsequently tested on unlabeled monitoring data from the same domain. TCA is used to project both datasets into a latent feature space where distribution discrepancies are minimized. This alignment allows knowledge gained from simulation to transfer effectively to field conditions despite modeling uncertainties. The study applied this framework to the Z24 bridge benchmark and confirmed that introducing domain. adaptation markedly improved classification consistency and reduced the dependence on model calibration

Martakis et al. [93] developed a hybrid framework that fuses damage-sensitive features with DA to improve damage classification robustness in real buildings. Numerical simulations are first employed to generate labeled acceleration data, which are then combined with limited field measurements. A convolutional neural network extracts key features, and a DA module based on maximum mean discrepancy ensures feature alignment across both domains. The framework prioritizes features that retain physical interpretability, focusing on those directly linked to modal and stiffness variations. Experimental results demonstrated that this hybrid design significantly improved model transferability from simulation to real buildings, achieving reliable classification under diverse environmental conditions.

Dynamic distribution alignment adapts to structural diversity such as span differences. Song et al. [83] developed a dynamic distribution adaptation network to perform bridge damage detection across structures of different spans without requiring any target labels. The model comprises two main components: a convolutional feature extractor that captures spatial and temporal patterns from vibration data, and a distribution alignment module that dynamically calibrates probability distributions between the source and target domains. This adaptive mechanism allows the model to handle differences in modal frequencies and stiffness distributions caused by span variations. Through numerical simulations of simply supported box-girder bridges, features extracted from a 32 m span structure were successfully transferred to a 24 m span target bridge. The method achieved precise localization and quantification of damage with no labeled target data, outperforming both static DA and conventional deep networks in terms of convergence speed and classification accuracy. The results highlight the effectiveness of dynamically aligning data distributions for generalizing across structurally diverse bridge populations.

Yuan et al. [94] introduced an unsupervised cross-domain framework to detect and localize damage in vibration isolators of metro floating-slab tracks. Their approach integrates adversarial training with statistical feature alignment to learn representations that are simultaneously domain-invariant and damage-sensitive. A convolutional neural network first extracts dynamic vibration features from both numerical simulations and field measurements, followed by a domain discriminator trained through adversarial learning to enforce indistinguishability between the two feature distributions. Simultaneously, a maximum mean discrepancy term ensures quantitative statistical consistency. The model demonstrated strong generalization when applied to real metro line data, accurately identifying degraded isolators and their positions even in the absence of labeled target samples. This study confirmed that combining adversarial alignment and unsupervised DA offers a powerful solution for condition monitoring of large-scale transportation systems.

Cao et al. [95] proposed an unsupervised DA strategy for post-earthquake damage identification in high arch dams, addressing the scarcity of labeled seismic data and the influence of varying reservoir levels. The framework employs a denoising sparse autoencoder enhanced with contractive constraints to extract stable damage features from dynamic response signals. To transfer knowledge between pre-earthquake and post-earthquake domains, a maximum mean discrepancy regularization term minimizes the distribution shift in the latent representation. The model is trained entirely on simulated vibration data and tested on measured responses under fluctuating water loads. Results demonstrated that the approach effectively identified stiffness degradation patterns induced by seismic excitation, maintaining accuracy across different reservoir conditions. The work highlights the potential of unsupervised deep transfer learning in hydraulic structures where environmental and operational variability heavily affect dynamic behavior.

Quqa et al. [96] introduced a transfer learning-based approach for detecting cracks in structures coated with conductive nanocomposite films using electrical impedance tomography (EIT). The method uses synthetic impedance maps generated by finite element simulations as the labeled source data and experimental impedance measurements from real specimens as the target. TCA is applied to align the feature spaces of the two domains, allowing the classifier trained on simulated data to correctly identify crack locations in real tests. By embedding the governing equations of EIT within the adaptation process, the approach maintains physical consistency between simulated conductivity distributions and measured electrical responses, enabling accurate crack identification with minimal calibration.

Collectively, these studies demonstrate the continued evolution of simulation-to-real transfer techniques that incorporate both physics-based and data-driven principles. Dynamic distribution alignment enhances feature adaptability across bridges with geometric variation, adversarial and unsupervised adaptation enable condition monitoring under unlabeled field data, and hybrid Bayesian updating further mitigates model uncertainty. Together they provide a cohesive direction for developing generalizable, label-efficient, and physically interpretable damage detection frameworks applicable to real-world civil structures.

### 3.4.2 Synthetic-to-real transfer

The increasing integration of generative and DA techniques has significantly advanced the simulation-to-real transfer paradigm in SHM. This subsection focuses on recent methods that utilize synthetic or simulated data to bridge the gap between computational models and real-world measurements. These methods leverage adversarial learning, semantic feature extraction, and self-supervised adaptation to achieve knowledge transfer under label-free or limited-data conditions, providing an efficient alternative to costly experimental datasets.

Ge et al. [114] developed a physics-informed and self-attention-enhanced generative adversarial learning framework for unsupervised DA between numerical simulations and field-measured structural responses. Their model modifies the traditional CycleGAN by incorporating physical constraints derived from linear dynamic equations and transformer-based attention modules within both the generator and discriminator. These enhancements enable more accurate translation of acceleration responses between simulated and experimental domains while preserving modal and frequency-domain properties. Validation through tests on a steel beam and a large-scale bridge structure confirmed that the approach not only reduced simulation-to-measurement discrepancies but also achieved faithful reproduction of dynamic features, thus improving the reliability of simulation-derived health monitoring models.

When damaged measurements are scarce, undamaged-to-damaged translation offers a practical route to synthesize condition data. Luleci et al. [121] employed a Cycle-Consistent Generative Adversarial Network to perform undamaged-to-damaged domain translation for vibration-based SHM. The framework learns bidirectional mappings between structural states by training generators and discriminators to synthesize realistic vibration responses under damaged conditions, given only undamaged measurements. Through experiments on a steel grandstand, the network successfully produced transformed acceleration signals that reflected changes in modal energy and frequency shifts caused by damage, thereby enabling condition classification without explicit damage data. The study validated the capacity of generative models to create physically meaningful signal transformations, facilitating the extension of SHM applications to scenarios with limited or absent damaged-state data.

Complementarily, zero-shot semantic features enable transfer without any target labels. Wang et al. [72] proposed a simulation-derived semantic feature framework that enables zero-shot DA for rail fastener failure detection under unseen operational scenarios. The method extracts semantic features from simulated vehicle–track coupled dynamics using acceleration–displacement regression, ensuring that the learned features are physically interpretable and transferable. By integrating DA and contrastive learning, the framework aligns semantic representations from simulated and real track responses, enabling the synthesis of realistic failure features without labeled target data. The model achieved over 99% accuracy for detecting multiple fastener loosening levels using track acceleration data and 92% accuracy using vehicle-mounted sensors. This study demonstrated the potential of physics-consistent semantic representations for transferring diagnostic knowledge across domains in railway systems.

Self-supervised pretraining further learns domain-agnostic structure-sensitive representations. Xu et al. [73] proposed a self-supervised DA framework that bridges numerical simulations and experimental data for concrete damage classification. The method combines multi-channel convolutional encoders with an adversarial alignment module that enforces statistical similarity between synthetic and real vibration signals. By employing self-supervised contrastive pretraining, the model learns intrinsic representations of damage-related features independent of domain-specific noise. Applied to concrete beam datasets, the framework achieved accurate damage localization and classification without labeled experimental data, highlighting that self-supervision can enhance simulation-to-real transfer effectiveness in vibration-based SHM.

Xiong et al. [84] developed a multi-channel 1D-CNN integrated with deep autoencoder-based DA (DAE-DA) for zero-shot seismic damage diagnosis. Their network translates unseen target-domain vibration signals into latent representations learned solely from source data, allowing the model to generalize without any target samples during training. The proposed pipeline consists of a tailored vibration data preprocessor, a DAE-based domain alignment module, and a multi-channel CNN that fuses multi-sensor features for multiclass damage quantification. Validation using both ASCE benchmark models and a three-story RC frame tested on a shake table showed that the model achieves high prediction accuracy and low false-negative rates, confirming its ability to transfer knowledge effectively between simulated and experimental domains under a rigorous zero-shot setting.

Soleimani-Babakamali et al. [122] proposed a zero-shot transfer learning framework for SHM that integrates GANs with spectral feature mapping to achieve damage detection across heterogeneous structures without using any labeled target-domain data. The framework differentiates between undamaged and damaged conditions in the source domain and employs a DA module to transfer this knowledge to the target domain using only baseline signals. Through adversarial training, the generator learns to align the spectral distributions between source and target frequency domains, enabling zero-shot damage inference. The approach was validated using three distinct experimental datasets, the Z24 Bridge, the Yellow Frame, and the Qatar University Grandstand Simulator, and achieved AUC scores up to 0.95 and F1 scores exceeding 0.97, demonstrating strong generalization and scalability for large-scale SHM applications.

Finally, adversarial alignment with auxiliary discrepancy terms stabilizes transfer to structurally diverse bridges. Nie et al. [115] introduced an adversarial-based transfer learning framework for vibration-based bridge damage detection, emphasizing adaptation between numerically simulated and real bridge response domains. The method utilizes a feature extractor trained via adversarial optimization to achieve domain-invariant representations while maintaining sensitivity to damage-related features. By employing MMD as an auxiliary constraint, the model minimizes distributional shifts between simulated and measured vibration data. Applied to bridge datasets with varying boundary conditions and span lengths, the approach achieved robust detection of multiple damage levels and reduced false alarms compared with conventional CNNs and statistical distance-based DA models, showcasing the potential of adversarial alignment for cross-domain bridge diagnostics.

The recent advancements in simulation-to-real DA have significantly enhanced the capacity of SHM systems to operate without labeled experimental data. These studies collectively emphasize the effectiveness of integrating generative, adversarial, and autoencoder-based architectures in reducing the domain discrepancy between numerical simulations and real-world measurements. By leveraging spectral mapping, latent-space alignment, and adversarial optimization, these approaches enable knowledge transfer from richly simulated source domains to structurally diverse target domains, facilitating accurate and reliable damage detection even in zero-shot conditions. Such developments demonstrate that simulation-derived data, when properly adapted through domain-invariant feature learning, can provide a powerful foundation for large-scale, label-free monitoring frameworks in civil infrastructure applications.

# 4. Key challenges and future directions

While domain adaptation has opened new frontiers in SHM, several challenges remain to be addressed before its full potential can be realized. This section highlights the major obstacles that researchers and practitioners continue to face and outlines future directions emerging from recent studies. The discussion focuses on issues such as negative transfer and domain similarity, the scarcity and quality of labeled data, the interpretability and reliability of DA models, and advanced developments including multi-source adaptation, federated frameworks, and lifelong learning strategies for SHM. Each challenge represents an area where current approaches show promise but still fall short of achieving generalizable, physically consistent, and certifiable solutions. Finally, this section summarizes ongoing research trends that indicate how DA in SHM may evolve toward more robust and scalable applications in the coming years.

## 4.1 Negative transfer and domain similarity

Although DA has demonstrated strong potential in enhancing the transferability of SHM models, negative transfer remains one of the most critical obstacles to achieving robust and generalizable performance. Negative transfer occurs when knowledge learned from one domain deteriorates model accuracy when applied to another, often because the underlying dynamic characteristics or boundary conditions differ substantially. Early DA methods primarily aimed to minimize statistical distance between domains by aligning global distributions of vibration features, yet such alignment can obscure or distort the physically meaningful information required for identifying damage. The lack of a clear and measurable notion of domain similarity further complicates this issue. Structural differences in stiffness, modal coupling, or environmental excitation can all cause feature spaces to diverge, leading to unstable or even misleading adaptation outcomes [107, 109].

A key challenge therefore lies in defining and quantifying similarity between domains so that adaptation occurs only when meaningful physical correspondence exists. Existing metrics such as maximum mean discrepancy or correlation alignment measure feature statistics but fail to reflect mechanical consistency or modal proximity. Future research should explore hybrid similarity measures that jointly encode statistical and physics-based properties, combining frequency deviations, modal assurance criteria, and stiffness distributions into an interpretable similarity index. Such an indicator could act as a pre-adaptation filter, assessing whether the knowledge of a source structure is relevant to the target and preventing unnecessary or harmful alignment. Moreover, embedding these similarity measures into the DA training objective would allow the model to dynamically regulate the degree of adaptation according to physical compatibility, thereby reducing the risk of negative transfer [101, 102].

Another fundamental issue concerns the unequal or partial overlap between domains in practical SHM applications. Different bridges or buildings rarely share identical sensor configurations, material properties, or damage types. Conventional global alignment assumes complete overlap in label and feature space, which may force physically unrelated samples to align artificially, leading to degraded classification boundaries. Recent studies indicate that selective or partial DA can substantially mitigate this problem, where adaptation is performed only on transferable subspaces or specific frequency bands. Techniques such as entropy-based weighting or attention-driven sub-domain alignment allow the model to focus on physically relevant signals while suppressing low-confidence samples. Future work should investigate adaptive alignment frameworks that dynamically identify transferable components during training, ensuring that knowledge migration preserves both damage sensitivity and interpretability [119, 123].

In summary, the dual challenges of negative transfer and inadequate domain similarity definition represent foundational barriers to effective DA in SHM. Addressing these issues requires physically interpretable similarity metrics, selective adaptation strategies, and adaptive source weighting schemes that prevent over-alignment and preserve damage sensitivity. The integration of statistical learning with physical reasoning is expected to drive the next generation of domain adaptation frameworks toward reliable and scalable deployment across structurally diverse civil infrastructures.

## 4.2 Lack of labeled data and synthetic data generation

A persistent obstacle for domain adaptation in SHM is the chronic scarcity of labeled damage data. Most civil structures spend the vast majority of their life in healthy states, field campaigns seldom capture diverse fault scenarios, and controlled damage tests are costly and limited in scope. Domain adaptation can reduce distribution shift between structures or environments, yet it does not create new labels and it cannot on its own expose the model to damage patterns that are absent from the training set. For this reason, research has increasingly coupled domain adaptation with data generation or label efficient pretraining so that models acquire richer damage sensitive representations before transfer. The central idea is to enlarge and diversify the exposure of the feature extractor while preserving physical plausibility, then adapt these features across domains with minimal supervision.

Generative modeling offers a practical route to supply the missing variability. Cycle consistent translation has been used to transform undamaged signals into damaged like responses, which expands the training distribution without requiring ground truth labels on the target domain. Physics informed generative adversarial learning further tightens this loop by constraining the generator with dynamic relations so that synthesized signals respect stiffness mass damping trends and preserve modal structure while injecting damage effects. When such synthetic waveforms are combined with domain adaptation, the feature extractor can learn to separate domain shift from damage variation and to transfer this separation to the target bridge or building. Studies that translate between numerical and field responses show that adversarial alignment and spectral mapping can reduce the simulation to measurement gap and enable zero shot diagnosis, indicating that synthetic data can be effective when it is both diverse and physically consistent [114, 121].

Self-supervised learning complements generation by exploiting the large volumes of unlabeled healthy monitoring data that are routinely collected. Pretext objectives such as reconstruction in time or frequency, masked segment prediction, or contrastive discrimination across operational regimes can teach a network to encode structural dynamics without labels. When followed by domain adaptation and a small number of target annotations, these pretrained encoders transfer more reliably because they already internalize invariant response patterns that are shared across sites and sensors. Results that pretrain on unlabeled measurements and then adapt to limited damaged data confirm that self supervision reduces overfitting to a specific sensor layout or excitation condition and stabilizes adaptation under environmental variability. Recent work further shows that making the pretext objective itself domain aware can align hidden representations before any labels are used, which lowers the burden on the final adaptation stage [72, 73].

Heterogeneous sensing is another practical impediment to label use. Bridges and buildings are monitored with mixtures of accelerometers, strain gauges, guided waves, and sometimes imaging or electrical impedance systems. Labeled damage examples may exist for one modality in a laboratory setting while the field asset relies on another. Cross modal domain adaptation addresses this mismatch by learning a shared latent space in which modalities that observe the same physics co locate even when their measurements are fundamentally different. In practice this means transferring features learned from labeled strain or guided wave data to unlabeled vibration or acoustic records on the target, while aligning class conditional structure so that damage sensitive cues remain separable after fusion. Multi channel adaptation on one dimensional signals has already shown that aligning conditional distributions across sensors improves population to one transfer; extending the same principle across modalities would unlock much broader reuse of scarce labels [101].

Collaboration across asset owners can also expand the effective source domain without moving raw data. Federated training shares model updates rather than signals, which respects privacy and operational constraints while still allowing a global feature extractor to see a much larger variety of operating conditions and benign variability. Embedding domain adaptation into this collaborative loop would let each participant specialize the global representation to its local domain while benefiting from the diversity contributed by others. The main technical issues are communication efficiency, bias from structurally dominant contributors, and the need to weight contributors according to structural similarity so that harmful updates are not propagated. Population based SHM provides a basis for such weighting by representing structures in a common space and regulating contribution by proximity, which in turn makes the federated adaptation process auditable for engineers [107].

Taken together, progress on synthetic generation, self-supervised representation learning, cross modal alignment, and collaborative training forms a coherent response to the label scarcity that motivates domain adaptation in SHM. The common thread is to expand the model's exposure to physically meaningful variability before and during adaptation, while keeping alignment guided by structure consistent cues rather than by purely statistical similarity. When physics informed synthesis populates plausible damage scenarios, when self supervision extracts invariant dynamics from unlabeled data, and when cross modal and collaborative schemes unlock labels that would otherwise be siloed, domain adaptation can operate with far fewer direct annotations on the target and still deliver reliable damage discrimination.

## 4.3 Interpretability and certification of domain adaptation Models

As DA models become increasingly adopted in SHM, their interpretability and certifiability emerge as critical prerequisites for practical deployment in safety-sensitive infrastructures. When a DA model indicates that damage exists at a specific location, engineers and asset managers must understand what structural features or sensor responses led to this conclusion. Unlike traditional signal-based indicators, DA systems integrate knowledge from other structures, simulations, or operational conditions, which introduces another layer of uncertainty regarding the reliability of their internal reasoning. Hence, developing interpretable and verifiable DA frameworks has become one of the central challenges for their acceptance in engineering practice [109].

Interpretability in DA primarily concerns the transparency of the feature transfer process. Most existing models rely on deep feature alignment through adversarial, statistical, or self-supervised objectives, which often yields abstract latent representations that lack explicit physical meaning. To foster trust, researchers are now exploring methods that make the adaptation process more transparent and observable. One approach is to employ attention-based mechanisms that reveal which portions of the vibration signal or which sensors contribute most to a model's decision. Visualization tools such as attention-weight maps or gradient-based saliency plots can indicate whether the DA model is focusing on frequency components or temporal segments known to be sensitive to stiffness degradation or crack initiation. When the emphasized features correspond to physically interpretable responses confidence in the model's reasoning increases. These approaches allow engineers to validate that adaptation does not simply fit statistical noise but captures meaningful mechanical behavior differences between domains [77, 112].

Beyond visualization, incorporating physical principles directly into the DA architecture provides a more fundamental path toward interpretability. Physics-informed domain adaptation integrates structural dynamics into network training, ensuring that transferred features and outputs remain consistent with governing equations or modal properties. For instance, instead of adapting raw vibration signals, models can operate on modal parameters such as natural frequencies, damping ratios, or transmissibility functions, which are inherently interpretable and describe meaningful changes in the system's dynamic characteristics. Enforcing equilibrium or compatibility relations in the latent feature space further constrains the learned representations to physically admissible behaviors. These hybrid frameworks that fuse data-driven and physics-based components have demonstrated greater reliability in identifying damage while maintaining interpretability, bridging the gap between black-box learning and physical reasoning [111, 114].

Certification and verification form another indispensable dimension for DA-based SHM. Since these models inform maintenance planning and safety assessments, engineers and regulators require demonstrable evidence that they behave reliably within prescribed operational ranges. Certification typically involves verifying that under nominal conditions corresponding to healthy states, the model consistently yields non-damage classifications, while under known or simulated damage scenarios, it detects anomalies with bounded uncertainty. Although challenging for neural architectures, DA can simplify this process by isolating the verification of the base model and the adaptation module. The base model can be validated with controlled laboratory data, while the adaptation procedure is verified using limited field samples or physics-consistent simulations. This layered approach ensures traceable and explainable verification, making DA models more defensible for engineering deployment [104].

Quantifying model confidence provides another pathway to certification. By incorporating probabilistic calibration or Bayesian updating within DA, predictions can be accompanied by uncertainty estimates that reflect both model and domain-shift variability. These uncertainty bounds allow engineers to determine when automated predictions are reliable and when additional inspection is required. For instance, physics-guided Bayesian DA has been shown to maintain transfer performance while quantifying epistemic uncertainty, producing results that are both accurate and trustworthy. Embedding such probabilistic reasoning into SHM workflows supports risk-informed decision-making and bridges the gap between algorithmic outputs and actionable maintenance judgments [92].

In summary, interpretability and certification form the foundation of trustworthy domain adaptation in SHM. Integrating physics-informed constraints, uncertainty quantification, and explainable mechanisms provides a realistic route toward DA systems that are both accurate and transparent. As future research refines these elements, domain-adaptive SHM frameworks will evolve into verifiable, auditable, and certifiable systems, allowing knowledge transfer across structures to be performed with the same rigor and accountability expected in traditional engineering analysis.

## 4.4 Multi-source, multi-task, and lifelong domain adaptation

Traditional DA methods in SHM typically focus on transferring knowledge from a single source structure to a single target and solving a single diagnostic task, such as binary damage detection. However, practical SHM scenarios are inherently more complex: monitoring networks involve multiple structures of varying designs, materials, and operational conditions; diagnosis must often include not only detection but also localization, severity estimation, and prognosis; and structures evolve over time due to aging, repair, or environmental variation. Consequently, future DA research must advance toward frameworks capable of integrating information from multiple domains, handling multiple interrelated objectives, and continuously adapting as new data arrive [114].

The first direction toward this goal lies in expanding DA from single- to multi-source configurations. Real-world infrastructures rarely exist in isolation, and an abundance of potentially useful data from similar bridges, turbines, or buildings often remains underutilized. MSDA seeks to exploit this diversity by integrating knowledge from several related sources, thereby enhancing generalization and reducing the risk of negative transfer. Instead of naively pooling all available data, MSDA algorithms assign adaptive weights to each source according to its relevance to the target domain. Attention-based fusion and mixture-of-experts architectures have proven effective for identifying which sources contribute most meaningfully to the target task, enabling models to leverage complementary information while down-weighting mismatched domains [112]. Recent studies demonstrate that incorporating multiple experimental or simulated datasets substantially improves transfer robustness, particularly under changing environmental or operational conditions. Nevertheless, MSDA introduces new difficulties such as resolving conflicting domain knowledge and managing the increased computational cost of aligning several distributions simultaneously. Ongoing research aims to develop scalable ensemble-based alignment frameworks and automatic source selection strategies to ensure that heterogeneous datasets collectively enhance, rather than compromise, adaptation quality. Future work should establish scalable MSDA benchmarks and develop adaptive weighting mechanisms guided by domain similarity metrics, enabling models to automatically evaluate and select relevant source domains.

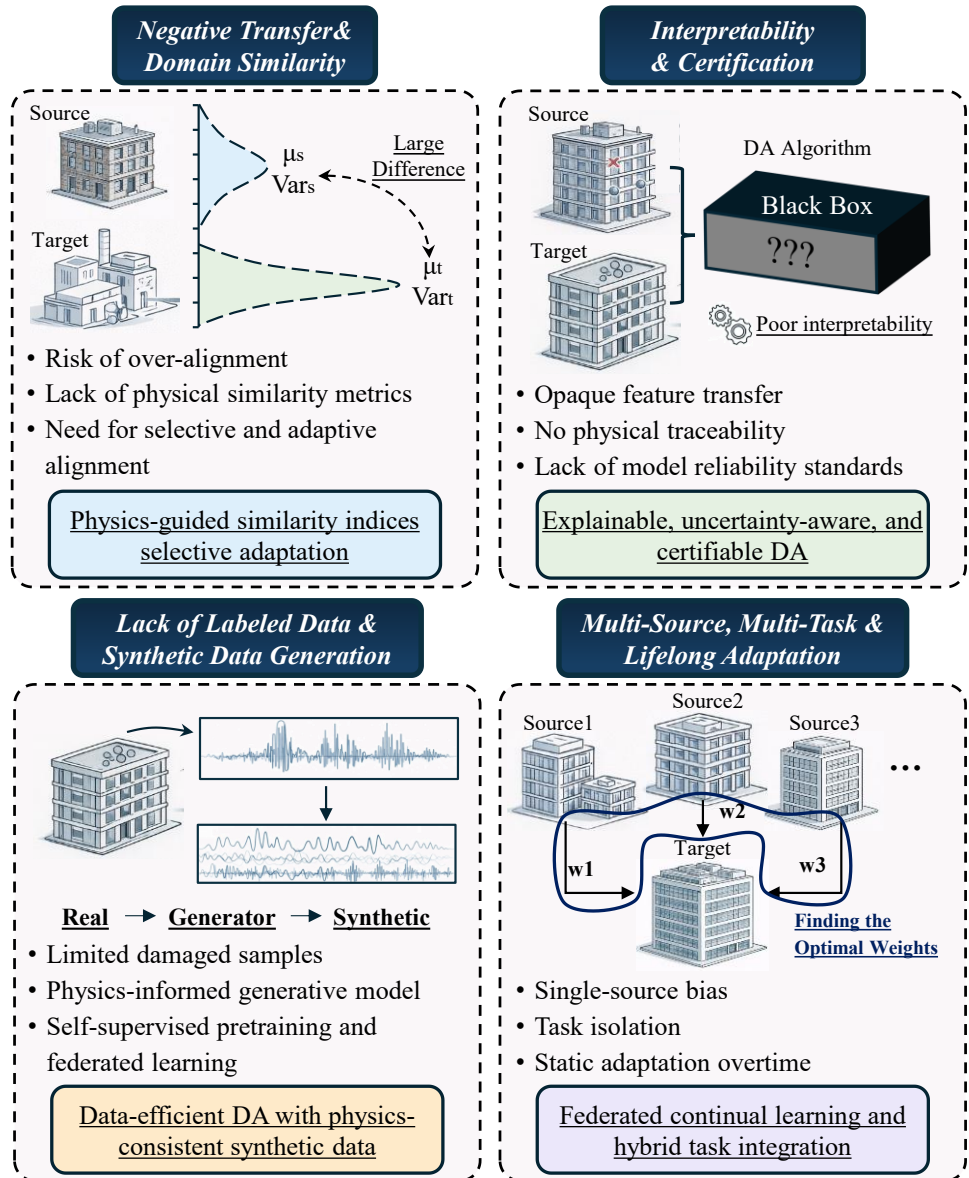

**Fig. 13.** Current challenges and future research directions.

Beyond integrating multiple data sources, extending DA to multi-task settings represents another critical step toward more comprehensive SHM intelligence. Real structures demand multidimensional diagnostics—identifying not just whether damage exists, but where it occurs, how severe it is, and how it may progress. Multi-task domain adaptation (MTDA) addresses this by sharing learned representations across several related objectives. A single model can jointly perform classification of damage type, regression of severity, and even prediction of temporal degradation trends, with each task reinforcing the others through shared latent features [118]. This shared representation enables the model to capture more complete and physically meaningful information about the structure's behavior, improving both generalization and interpretability. For example, features that are useful for localizing damage often highlight spatial vibration modes that also benefit global detection. By leveraging complementary supervision among tasks, MTDA reduces overfitting and enhances the robustness of adaptation under variable environmental conditions. Furthermore, embedding physical constraints and uncertainty quantification within MTDA frameworks can ensure that simultaneous predictions remain physically plausible, aligning data-driven inference with engineering principles. Future studies may explore hybrid task hierarchies where detection, localization, and prognosis are jointly optimized under shared physics-informed representations, ensuring that multi-task outputs remain consistent with structural dynamics.

The third essential direction concerns the temporal dimension of monitoring, developing models that adapt continuously as structures and their environments evolve. A static adaptation performed once between source and target domains cannot accommodate long-term changes such as gradual deterioration, repairs, or seasonal variations. Lifelong or continual domain adaptation (CDA) aims to overcome this limitation by enabling models to incrementally update their knowledge without losing prior information. Each new operating condition or time period can be viewed as a new domain, and small-scale updates can be performed periodically to maintain accuracy. Techniques such as memory replay, elastic weight consolidation, and regularized updating have been applied to prevent catastrophic forgetting, allowing the model to retain knowledge of past damage signatures while learning new ones [68]. In SHM applications, CDA has shown promise for tracking slow degradation trends and adjusting to environmental drift, thereby ensuring stable performance over extended operational periods. Integrating CDA into digital-twin-based SHM frameworks provides a natural implementation pathway, as digital twins continuously assimilate structural responses and update their underlying physics and data models in parallel. Developing benchmark datasets and standardized evaluation metrics for continual adaptation in SHM would greatly accelerate practical adoption, allowing models to evolve with structures under controlled and comparable settings.

Despite these advancements, several open challenges remain. Distinguishing between genuine structural deterioration and benign environmental fluctuations is nontrivial. An overadaptive model may misclassify seasonal effects as damage, while an underadaptive one may overlook early signs of distress. Future work should focus on adaptive drift detection and probabilistic thresholding schemes that allow the model to balance sensitivity with reliability. Moreover, continual adaptation must scale beyond individual assets. Federated learning concepts offer a promising avenue by enabling multiple structures to collaboratively improve a shared global model without exchanging raw data, thereby preserving privacy and promoting scalability [111]. In such a federated continual learning framework, knowledge gained from one structure can inform another, creating an interconnected SHM ecosystem where collective experience accelerates learning and enhances reliability.

Overall, the evolution of DA in SHM toward multi-source, multi-task, and lifelong paradigms marks a transition from isolated, one-time transfer to a dynamic, knowledge-sharing process, which is conceptually framed in Fig. 13. These developments will enable SHM systems to accumulate expertise across structures, tasks, and time, mirroring how human engineers learn from diverse experiences. By coupling DA with scalable fusion, multi-objective reasoning, and continual updating, future monitoring systems can achieve sustained accuracy, resilience, and autonomy, ultimately realizing a self-evolving infrastructure intelligence that improves with every new observation.

## 5. Conclusion

This review systematically examined the recent evolution of DA techniques for vibration-based SHM, emphasizing how DA bridges the persistent gap between laboratory simulations, numerical modeling, and real-world field measurements. Drawing on more than sixty representative studies, the paper consolidates the diverse methodological landscape of statistical, adversarial, self-supervised, and physics-informed learning frameworks, illustrating how these approaches have been progressively integrated to enhance knowledge transfer across structural types and sensing configurations. The synthesis reveals a clear trajectory in the field: data-driven SHM methods are transitioning from isolated and site-specific implementations toward transferable, physics-constrained frameworks capable of generalizing across different structures, sensors, and operational environments.

DA has shown considerable potential in enabling knowledge transfer between structures with different boundary conditions, materials, and geometries, as well as in mitigating the effects of environmental and operational variability. Statistical alignment methods, such as TCA, JDA, and LMMD, have improved cross-domain feature consistency by minimizing distributional shifts in the latent space, while adversarial learning has provided flexible mechanisms for high-dimensional feature matching through domain-confusion strategies. The incorporation of physics-based constraints has further enhanced interpretability and generalization, ensuring that transferred features remain consistent with structural mechanics. Collectively, these studies confirm that DA can substantially reduce data dependency, enhance model robustness, and promote scalable applications of intelligent SHM across diverse infrastructure systems.

Nevertheless, several key challenges remain before DA can be fully deployed in real engineering practice. The first major challenge concerns domain discrepancy and negative transfer. Although statistical and adversarial strategies can align distributions, they often ignore the underlying physics of vibration responses, leading to loss of physical interpretability and occasional instability during adaptation. A second challenge involves the scarcity of labeled damage data, which restricts supervised learning and limits the validation of complex DA architectures. While self-supervised and generative approaches have been explored to produce synthetic or augmented datasets, ensuring their physical fidelity remains an open issue. The third challenge relates to interpretability and certification. DA models, particularly deep neural ones, are typically viewed as black boxes; without transparent reasoning or traceable links to modal properties, they cannot be easily trusted in safety-critical infrastructure. Finally, the extension to multi-source, multi-task, and lifelong learning scenarios poses a frontier problem. Current studies mostly adopt single-source and single-task configurations, which cannot accommodate continuously evolving structural states, distributed sensing networks, or heterogeneous datasets accumulated over time.

Looking ahead, several research pathways are expected to shape the next stage of DA-based SHM. A primary direction is the development of physics-informed DA frameworks that integrate equilibrium relationships, modal properties, and stiffness–frequency correlations directly into learning objectives. Such integration will align latent features with structural dynamics, reducing reliance on purely statistical alignment and enabling models to generalize with physical interpretability. A second opportunity lies in synthetic-to-real transfer through generative modeling. Physics-constrained GANs and diffusion networks can create realistic vibration or damage scenarios consistent with governing equations, providing reliable data for pre-training and zero-shot adaptation when field labels are scarce. Equally important is the pursuit of interpretable and certifiable DA models. Future systems must incorporate explainable mechanisms to expose their decision logic and ensure accountability in safety-critical applications. Minimal validation on target structures, combined with probabilistic confidence estimation, could become a practical framework for certification. Moreover, multi-source, multi-task, and lifelong adaptation will be essential for scaling DA to real operational networks. Leveraging knowledge from multiple structures, sharing representations across related tasks, and continuously updating models through digital twins or federated schemes will support long-term, adaptive monitoring across infrastructure portfolios.

In summary, domain adaptation has transformed SHM from isolated case-specific modeling into a transferable, knowledge-driven paradigm that connects simulations, experiments, and in-service monitoring. Its continued progress depends on integrating physical laws, improving data realism, ensuring interpretability, and enabling continual learning across distributed structures. With these advancements, DA-empowered SHM systems are expected to evolve into intelligent, transparent, and adaptive platforms that support proactive maintenance and long-term resilience of civil infrastructure.

# 6. Declaration of conflicting interests

The authors declared no potential conflicts of interest with respect to the research, authorship, and/or publication of this article.

# 7. Funding

The author(s) received no financial support for the research, authorship, and/or publication of this article.

# 8. Data availability

Data will be made available on request.